\documentclass[a4paper]{article}

\usepackage{anysize}
\marginsize{2.3cm}{2.3cm}{2.3cm}{2.3cm}

\usepackage[table]{xcolor}
\usepackage{graphicx,amssymb,verbatim,epsf,amsmath,setspace,float,dsfont, soul}
\usepackage{tikz}
\usetikzlibrary{shapes,arrows,decorations.pathmorphing,backgrounds,positioning,fit} 
\usepackage[numbers, comma, round, sort&compress]{natbib}
\usepackage{amsmath,amsthm,afterpage}
\newcolumntype{C}[1]{>{\flushright\let\newline\\\arraybackslash\hspace{0pt}}m{#1}}

\makeatletter 
\renewcommand{\thefigure}{S\@arabic\c@figure}
\makeatother
\numberwithin{equation}{section}
\newcommand{\bi}{\begin{itemize}}
\newcommand{\ei}{\end{itemize}}
\definecolor{dr}{HTML}{CC0000}

\newcommand{\bcat}{$\beta$-catenin }
\newcommand{\etal}{{\em et al.} }
\newcommand{\y}{\checkmark}

\newtheorem{thm}{Theorem} 

\newtheorem{defn}[thm]{Definition}   
\newtheorem{remark}{Remark}   

\newcommand{\obs}{\mathrm{obs}}

\usepackage{array}
\newcolumntype{C}[1]{>{\centering\let\newline\\\arraybackslash\hspace{0pt}}m{#1}}

\begin{document}
\title{Parameter-free methods distinguish Wnt pathway models and guide design of experiments}
\author{Adam L MacLean$^{1}$, 
Zvi Rosen$^{2}$,
Helen M Byrne$^{1}$
and
Heather A Harrington$^{1}$ \\ \\
$^{1}${Mathematical Institute, University of Oxford, Andrew Wiles Building, Oxford, UK}
\\
$^{2}${Department of Mathematics, University of California, Berkeley, USA}
}
\maketitle

\begin{abstract}
The canonical Wnt signaling pathway, mediated by $\beta$-catenin, is crucially involved in development, adult stem cell tissue maintenance and a host of diseases including cancer. We  analyze existing mathematical models of Wnt and compare them to a new  Wnt signaling model that targets spatial localization, our aim is to distinguish between the models and distill biological insight from them. Using Bayesian methods we infer parameters for each model from mammalian Wnt signaling data and find that all models can fit this time course.  We appeal to algebraic methods (concepts from chemical reaction network theory and matroid theory) to analyze the models without recourse to specific parameter values. 
These approaches provide insight into aspects of Wnt regulation: the new model, via control of shuttling and degradation parameters, permits multiple stable steady-states corresponding to stem-like vs committed cell states in the differentiation hierarchy.
Our analysis also identifies groups of variables that should be measured to fully characterize and discriminate between competing models, and thus serves as a guide for performing minimal experiments for model comparison. 
\end{abstract}

\section*{Keywords} Experimental design | Bistability | Bayesian inference | Matroids | CRNT
\section*{Abbreviations} APC, adenomatous polyposis coli; CRNT, chemical reaction network theory; DC, destruction complex; Dsh, Dishevelled; GSK, glycogen synthase kinase; TCF, T-cell factor

\section{Significance Statement}
The canonical Wnt/\bcat signaling pathway is important for essential cellular functions such as development, homeostasis and is implicated in many diseases. We introduce a new mathematical model that focuses on \bcat degradation and protein shuttling between cellular compartments. We compare our model to others and show that all fit to time-dependent experimental data. To evade this parameter problem, 
we use algebraic methods and characterize model features that are independent of the choice of parameter values. We find that multiple responses to Wnt are feasible under certain conditions for the new model, but not for the others; moreover we provide dependencies between species (variables) that inform future experiments and model discrimination. We also highlight the wide applicability of these tools across problems in systems biology.


\section{Introduction}
The Wnt signaling pathway plays a key role in essential cellular processes ranging from proliferation and cell specification during development to adult stem cell maintenance and wound repair \cite{Logan:2004fz}. Dysfunction of Wnt signaling is implicated in many pathological conditions, including degenerative diseases and cancer \cite{Polakis:2000pa, Reya:2005js, Vermeulen:2010ve}. Despite many molecular advances, the pathway dynamics are still not well understood. Theoretical investigations of the Wnt/\bcat pathway serve as testbeds for working hypotheses \cite{Lee:2003bw, vanLeeuwen:2007ga, Schmitz:2013kn, Mirams:2010ha, Cho:2006jt, Goldbeter:2008do, Fletcher:2012cc, vanLeeuwen:2009ev}. 
\par
We focus on models of canonical Wnt pathway processes with the aim of elucidating mechanisms, predicting function, and identifying key pathway components in adult tissues, such as colonic crypts. 
We compare four pre-existing ordinary differential equation models \cite{Lee:2003bw, vanLeeuwen:2007ga, Schmitz:2013kn, Mirams:2010ha}, and find, using injectivity theory, that for any given conditions and parameter values, none of the models are capable of multiple cellular responses.
\par
In many tissues Wnt plays a crucial role in cell fate specification \cite{Reya:2005js}. At the base of colonic crypts, cells exist in a stem-like, proliferative phenotype in the presence of Wnt. As these cells' progeny move up the crypt axis they enter a Wnt-low environment and change fate (perhaps reversibly), becoming differentiated, specialized gut cells \cite{Philpott:2014ha}. In neuronal and endocrinal tissues, Wnt/\bcat data suggest cell fate plasticity under different environmental conditions \cite{Bryja:2007vm, Hannoush:2008ku}. Here, we introduce a new model motivated by experimental findings not described in previous models \cite{Li:2012jw, Habas:2005dq, Li:2001hc} in order to investigate bistable switching in the Wnt pathway. We find the new model to be capable of multiple cellular responses; furthermore, our parameter-free techniques identify that molecular shuttling (between cytoplasm and nucleus) and degradation together may serve as a possible mechanism for governing bistability in the pathway, corresponding to, for example, a committed cell state and a stem-like cell state.
\par
Comparison of models (and mechanisms) requires data; the type of comparison performed depends on the data at hand. If data show bistability (two distinct response states), then we could rule out all models that preclude bistability; however the converse is not true (a graded response may be compatible with all models). Experimental studies in \emph{Xenopus} extracts have been performed to validate a model of Wnt signaling \cite{Lee:2003bw}, with further pathway elucidation in \cite{Goentoro:2009dj, Hernandez:2012cw}; however the parameters identified in these studies may differ markedly from those involved in mammalian Wnt signaling \cite{Tan:2012hw,Tan:2014ii}. With the aim of discriminating between models, we present the five Wnt models  under a unifying framework, 
with standardized notation to facilitate comparison. We fit parameters to recently published mammalian \bcat signaling time course data using Bayesian inference \cite{Tan:2014ii} and find that all of the studied models can describe the data well, demonstrating that additional data are required to compare models.
\par
In order to determine which sets of protein species should be measured for carrying out data/model comparison, we introduce matroid theory to systems biology. A matroid is a combinatorial structure from mathematics, and in our case, it provides all of the steady-state invariants \cite{Manrai:2008kb,Harrington:2012us} that have minimal sets of variables. The algebraic matroid associated to the steady-state ideal determines specific sets of species that should be measured to perform model discrimination without knowledge of parameter values. We demonstrate this parameter-free analysis for two Wnt models.
 \par
In the next section, we introduce the previous models and new shuttle model. We perform injectivity/multistability analysis and classify the shuttle model as the only one capable of multistability. Next we infer the parameters of five competing models for time-course \bcat data, revealing that all the models fit the data. Finally we introduce algebraic matroids to inform experimental design for discriminating between models and data.

\section{Models}
Over the past decade, Lee {\em et al.}'s seminal model of canonical Wnt signaling \cite{Lee:2003bw} has spawned many variants. Briefly, the underlying biology of the pathway that these models describe is as follows \cite{LloydLewis:2013ft}: 
Wnt binds to cell-surface receptors that transduce a signal via a multi-step process involving Dishevelled (Dsh) to the so-called destruction complex (DC), which contains forms of Axin, adenomatous polyposis coli (APC) and glycogen synthase kinase 3 (GSK-3). In the absence of a Wnt signal, the DC actively degrades \bcat -- which is being continually synthesized in the cell -- by phosphorylating it and marking it for proteasomal degradation. Following Wnt stimulation,  degradation of \bcat is inhibited through phosphorylation of DC members, leading to accumulation in the cytoplasm of free $\beta$-catenin, which is able to translocate to the nucleus where it can form a complex with T-cell factor (TCF) and lymphoid-enhancing factor proteins and influence the transcription of target genes. These are cell-type specific, although genes controlling self-renewal and proliferation are commonly regulated across many cell types \cite{Logan:2004fz}. 
\par
\begin{figure}[h!]
\includegraphics{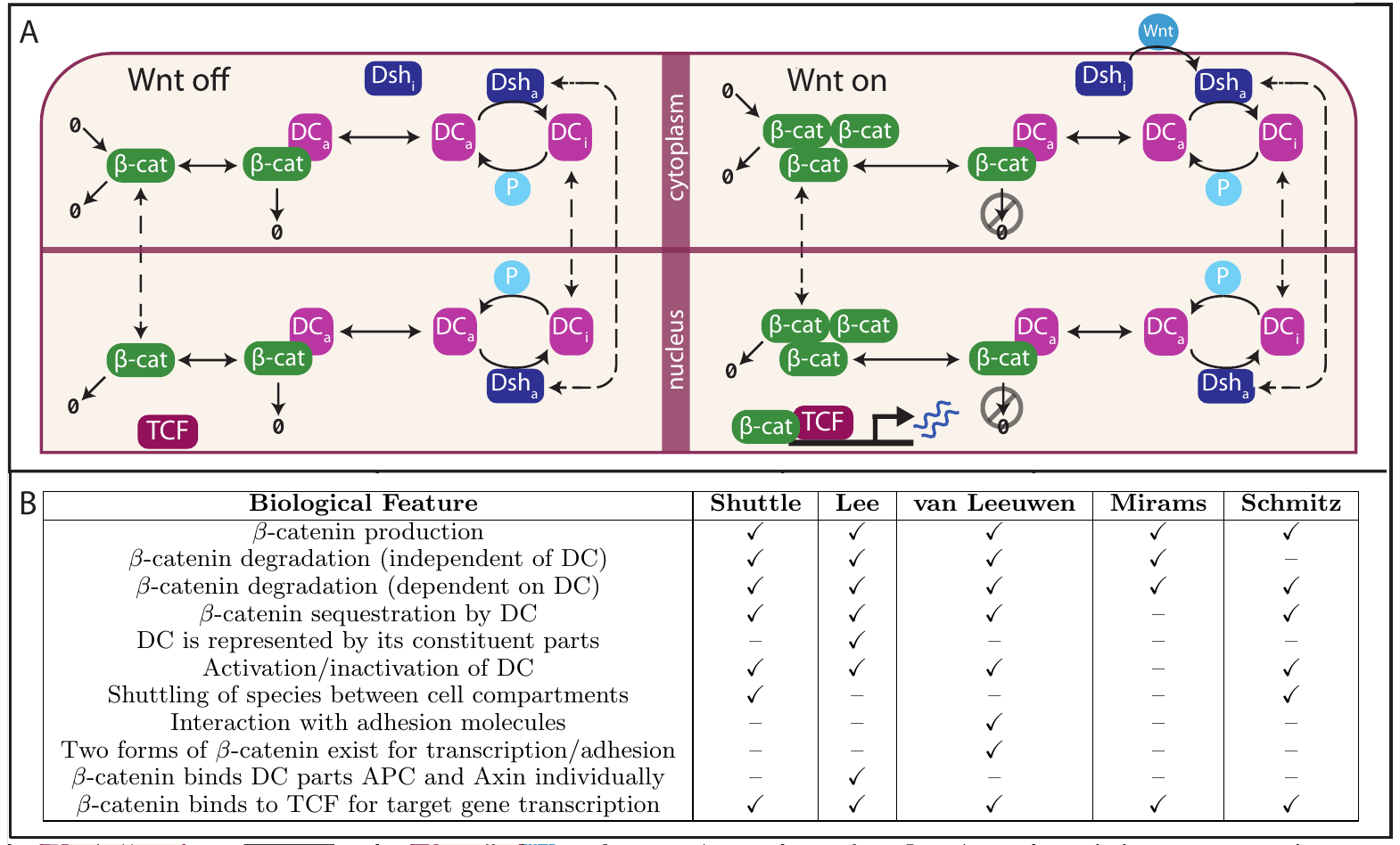}
\caption{Comparison of models of Wnt signaling. {\bf (A)} Schematic of the new shuttle model in absence (left) and presence (right) of a Wnt stimulus. DC: destruction complex, and Dishevelled (Dsh) exist in inactive (i) and active (a) forms.  P: phosphatase. For a full description of the reactions specifying this model, see SI. {\bf (B)} Comparison of models: the shuttle model is compared with four others from literature \cite{Lee:2003bw, vanLeeuwen:2007ga, Schmitz:2013kn, Mirams:2010ha} based on the features that are present/absent in each model.
}
\label{fig-newmodel}
\end{figure}

We include the core processes as well as the following in the proposed shuttle model (Figure \ref{fig-newmodel}A and SI): 
\begin{enumerate}
\item spatial localization of shuttling components Axin, APC, GSK-3, and Dsh, the importance of which has been reported for each species 
\cite{FrancaKoh:2002ia, Wiechens:2004fn, Cong:2004ew, Henderson:2002dl,Itoh:2005kg, Habas:2005dq}; 
\item an alternative degradation mechanism whereby \bcat is degraded while still bound to active DC and sequestered but not degraded by inactive DC \cite{Li:2012jw};
\item catalysis of the reverse reaction by Phosphatase (P) that converts DC from inactive to active form by dephosphorylating members of the DC  \cite{Luo:2007hm, Li:2001hc, Strovel:2000hm}. 
\end{enumerate}
\par
The behavior of four other published models are analyzed and compared with that of the shuttle model \citep{Lee:2003bw, vanLeeuwen:2007ga, Mirams:2010ha, Schmitz:2013kn}. Figure \ref{fig-newmodel}B summarizes the distinguishing qualitative features of each model; full model descriptions, using a standardized notation (that differs from the authors' originals) are summarized by a composite model in the Supporting Information.

\section{Results and discussion}
Wnt signaling interaction networks are polynomial systems whose steady state solutions are defined by sets of algebraic equations for the species' concentrations; this opens up avenues for parameter-free analysis, as we show here. 

\subsection{Parameter-free analysis of Wnt models I: Multistability}
We are interested in determining whether or not a given model can produce multiple positive stable responses (states). Standard approaches from dynamical systems (e.g. bifurcation and singularity theory  \cite{Guckenheimer1983,Kuznetsov98,Golubitsky1985}) are useful for small systems or when we have knowledge of the parameters, however, for systems of more than a few free variables (the shuttle model has 19 species and 31 parameters), such approaches become infeasible. To overcome this, we apply theory developed for chemical reactions to Wnt pathway  models; this is particularly helpful for determining whether multiple states are possible without the need for parameter values or sampling. There are various conventions in chemical reaction network theory (CRNT) for describing the number of positive steady states; we use the following terminology: 
\begin{enumerate} 
\item \emph{Injective}: implies at most one steady state. 
\item \emph{Multistationarity}: capacity for multiple steady states. 
\item \emph{Multistability}: capacity for multiple stable (accessible) steady states. \end{enumerate} 
\par
We test the injectivity of each model following graph-theoretic or Jacobian-based approaches used in CRNT \citep{Craciun:2005ub, Craciun:2006ki,craciun:pnasu:2006, Feliu:2012ft, Feliu:2013cf}. We find that only the Schmitz \etal and the shuttle model fail injectivity  and exhibit multistationarity. Further analysis reveals that the Schmitz \etal model is capable of at most two steady-states, only one of which is stable (see SI for proofs).  Whereas all of the previous models possess at most one positive stable steady state for any choice of the parameter values and conserved species' concentrations, the shuttle model has the capacity for multistationarity and multistability.  We find that when three or more species shuttle  (e.g. Dsh, inactive DC, and $\beta$-catenin), the model exhibits two stable states; we proceed to analyze this version of the shuttle model. Previous mathematical studies have proven shuttling across compartments is a mechanism for multistability \cite{Bhalla:2011wd,Harrington:2013gp}.
\par
\begin{figure}[h!]
\includegraphics{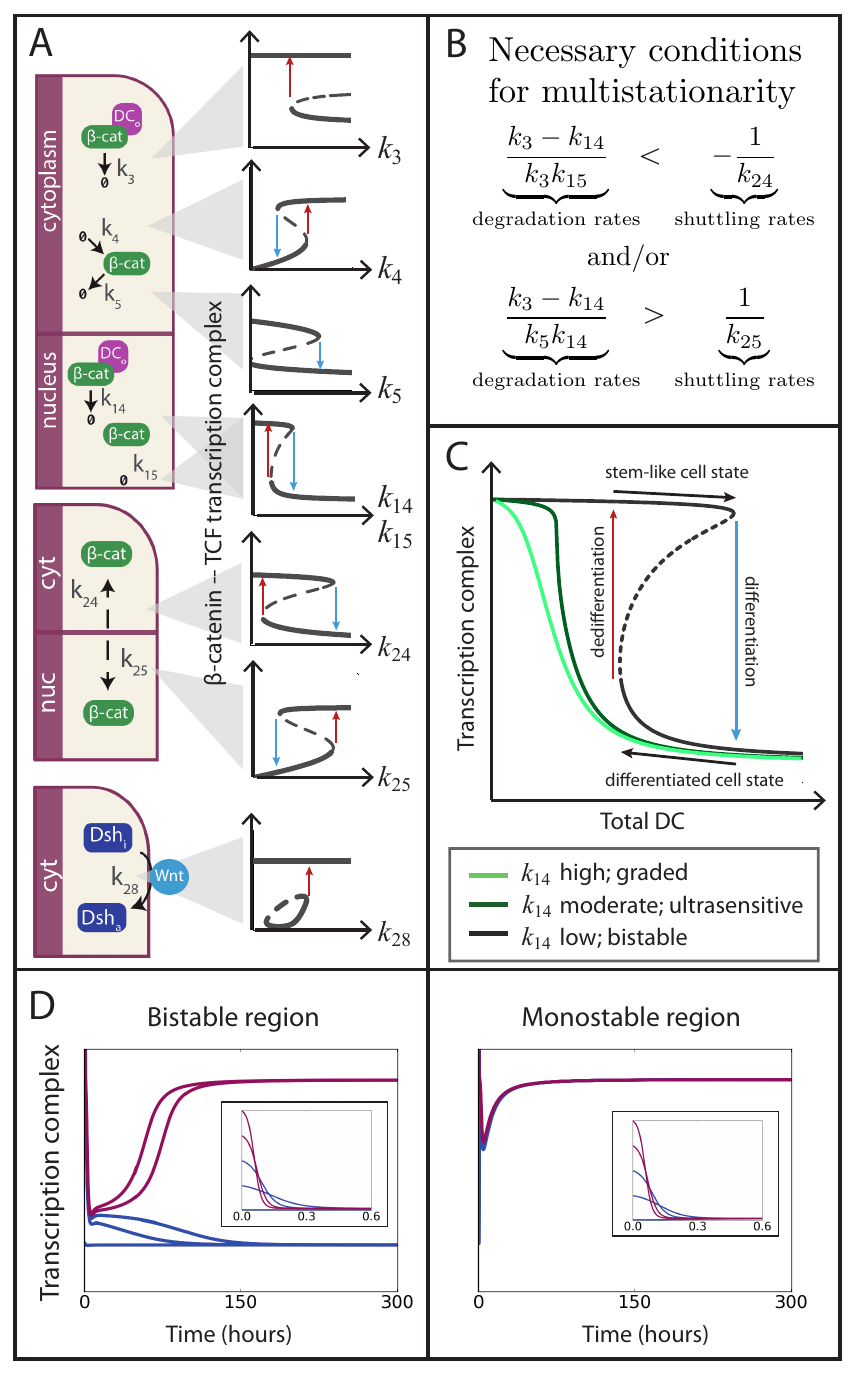}
\caption{
Bistability of shuttle model. {\bf (A)} Bifurcations diagrams as a proof-of-concept for feasible shuttle model behavior. Reversible and irreversible behaviors are observed; however could change if parameter values are known. In each case the high state of target gene transcription corresponds to a stem-like state and the low state corresponds to a differentiated cell state. {\bf (B)} Two of the eight necessary conditions for multistationarity of the shuttle model. {\bf(C)} Model exhibits different behaviors as degradation parameter $k_{14}$ is varied: for low values, bistability; for moderate values, switch-like (ultrasensitive) response; and for high values, graded response. {\bf (D)} Simulated trajectories for target gene transcription from five different initial conditions. In the bistable region we see two steady states reached; in the monostable region only the high (stem-like) state can be reached. Note that initial behavior in each region is similar: it is important to simulate for long enough to recover these differences in behavior.
}
\label{fig-bifs}
\end{figure}

For each model, we have a minimal collection $f_i$ of polynomial relations (including conservation laws) and $x_j$ of species. The model is injective if the determinant of the {\em Jacobian matrix} ($\partial f_i/\partial x_j$, which is polynomial in the parameters), has all positive or all negative coefficients: this ensures that the determinant is non-vanishing in the positive orthant where all feasible parameters lie. Although the shuttle model fails injectivity, we can find a set of eight sufficient conditions on the parameters (inequalities defining a semi-algebraic set contained in the injective region of parameter space) that precludes multistationarity (see SI). Negating these gives necessary, but not sufficient, conditions on the parameters for multistationarity, which depend explicitly on the degradation and shuttling rate constants (Figure \ref{fig-bifs}B). Since the parameter values are unknown, we compute an illustrative bifurcation diagram for a specific parameter set satisfying these necessary conditions. These diagrams are similar to dose-response curves in an experimental setting, and they demonstrate different behavior as the shuttling and degradation rate constants are varied (Figure \ref{fig-bifs}A,C). 
\par
Within a certain parameter region, we can either observe a graded response or switching and hysteresis behavior (Figure \ref{fig-bifs}C). The hysteresis loop shown by the black (stable state), blue (switch to committed state) and red (threshold switch to stem-like state) arrows enables switching at different thresholds between two stable steady states. While over short timescales bi/monostable behavior is indistinguishable (Figure \ref{fig-bifs}D), at steady state these differences emerge. In the bistable regime the low level of gene transcription is associated with a committed cell state and the high level with a stem-like phenotype over long time periods (Figure \ref{fig-bifs}D). As the value of a parameter, for example \bcat shuttling into the cytoplasm ($k_{25}$), decreases below a threshold, the bifurcation diagram predicts that cells will differentiate. If the shuttling rate was adequately increased, according to the diagram at these particular parameters, these cells would dedifferentiate to a stem-like state. If the parameter regime were known, bifurcation analysis and singularity analysis could also predict parameters governing reversible and irreversible behavior (e.g. $k_5$, irreversible in Figure~\ref{fig-bifs}A). 
\par
If qualitative data showed a clear bistable switch, then the shuttle model would be the best model. However, given quantitative rather than qualitative data, how can we compare models?

\subsection{Wnt model comparison via parameter inference}
Where competing models describe the same biological processes, one can perform parameter inference or model selection; such methods have been applied to a variety of problems in systems biology, ranging from cancer modeling to population genetics  \cite{MacLean:2014pn, Jiang:2009dp}. 
\par
\begin{figure}[h!]
\begin{center}
\includegraphics{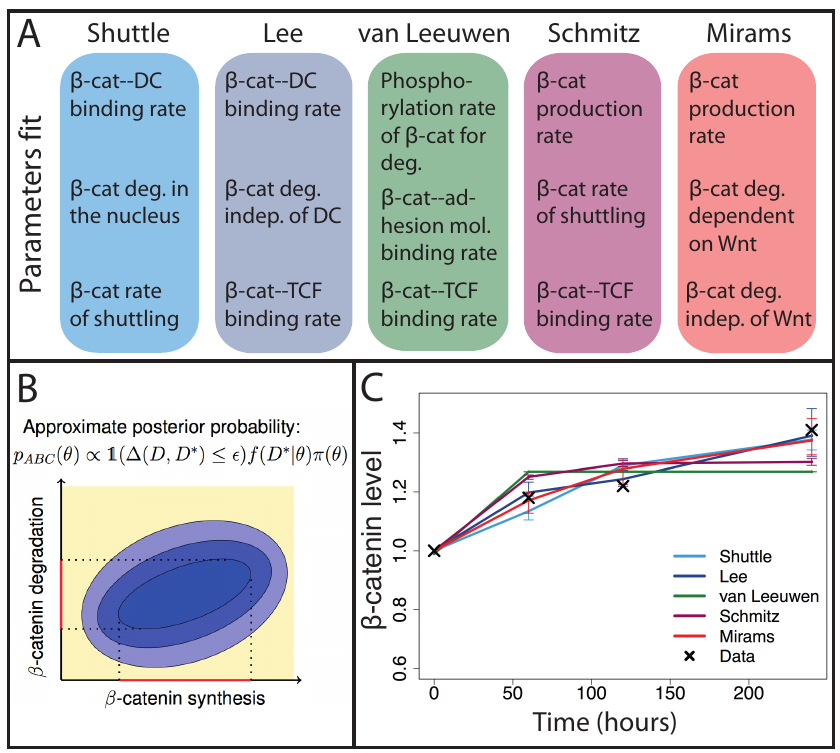}
\caption{
Bayesian parameter inference of Wnt signaling models. {\bf (A)} Description of the parameters that are inferred for each of the models used to fit the data describing \bcat dynamics following a Wnt stimulus. {\bf (B)} Depiction of the posterior probability distribution used for approximate Bayesian computation. The sequentially decreasing region of probability (blue ovals) defines the joint space of two parameters. Here we show synthesis and degradation rates that are a subset of the full parameter space. This is applicable to any of the Wnt models considered. {\bf (C)} Fits to the data simulated from the posterior distribution for each model (1000 particles simulated). Bars represent the 5\% and 95\% intervals. 
}
\label{fig-inference}
\end{center}
\end{figure}

Inferring parameters from data via Bayesian analysis provides the posterior probability distribution over the parameters, from which more information can be gleaned than by point estimates alone. In Figure \ref{fig-inference}B, we demonstrate the Bayesian inference procedure by considering a 2D subset of the parameter space (rates of \bcat synthesis and \bcat degradation). The panel shows how over successive iterations we can home in on the most probable region of parameter space given the data.
\par
Each of the Wnt models has a different number of parameters.  In an attempt to compare the models fairly and to reduce the size of the parameter space that we are searching, for each model we choose to fix all of the parameters (at estimated or arbitrary values) except for three. These three are allowed to vary and are used to fit the model to the Wnt/\bcat time course data recently published in \cite{Tan:2014ii}; see Methods for details. We chose the free parameters based on their point of influence on the pathway, targeting parameters with direct or near-direct influence on \bcat dynamics (Figure~\ref{fig-inference}A) and the fits we obtained after performing inference are shown in Figure~\ref{fig-inference}C. We see that even with only three degrees of freedom, good parameter fits are obtained for all of the models. Studying the posterior for each model reveals relationships between parameters: high \bcat production and low \bcat degradation rates are favored across models; but $\beta$-catenin-TCF binding rates vary considerably between models.  
\par
The disparity between model complexity and data availability prevents us from choosing between models based on model selection analysis. The problem could be addressed by simplifying models or collecting more data (additionally, experimental design influences Bayesian model selection results \cite{Silk:2014en}). Here, we proceed to use parameter-free methods to help guide experiments for model discrimination. 

\subsection{Parameter-free analysis of Wnt models II: Matroids}
Instead of classifying the feasible behaviors of the {\em whole} system, we can use the finer structure of a model to derive relations in each {\em part} of the system, for example, the concentrations of species in a chemical reaction network at steady-state. The matroid of a model is a list of \begin{enumerate} 
\item the subsets of species that are related, and \item the subsets of species whose concentrations are unrelated. \end{enumerate} 
A matroid is a set with a notion of independence for its subsets. The classic example of a matroid is an arrangement of vectors. Suppose $v_1 = (1,0,0), v_2 = (0,1,0), v_3 = (0,0,1), v_4 = (1,1,0)$, as in Figure 4A. A set of vectors is called \emph{dependent} if it is linearly dependent, i.e. there is a set of scalars, not all zero, such that multiplying by the vectors and adding them together results in the zero vector. Here, $\{v_1,v_2,v_4\}$ is dependent because $v_1 + v_2 - v_4 = (0,0,0)$. If no such relation exists, the set is called \emph{independent}. Any set of size three or smaller excluding 1-2-4 is independent (Fig 4A). The matroid does not remember the coordinates of the vectors, only whether any subset is dependent or independent.
The matroid construction in our discussion uses algebraic independence instead of linear independence. Explicitly, if there is a polynomial relationship that a collection of species satisfies at steady-state, they are considered a dependent set.  Note that this only considers relationships at steady-state, where the possible species concentrations describe an algebraic variety.
The independent and dependent sets of molecular players in each model may help compare models, guide experiments, and possibly reject models as described throughout this section.

\par
\begin{figure}[h!]
\begin{center}
\includegraphics{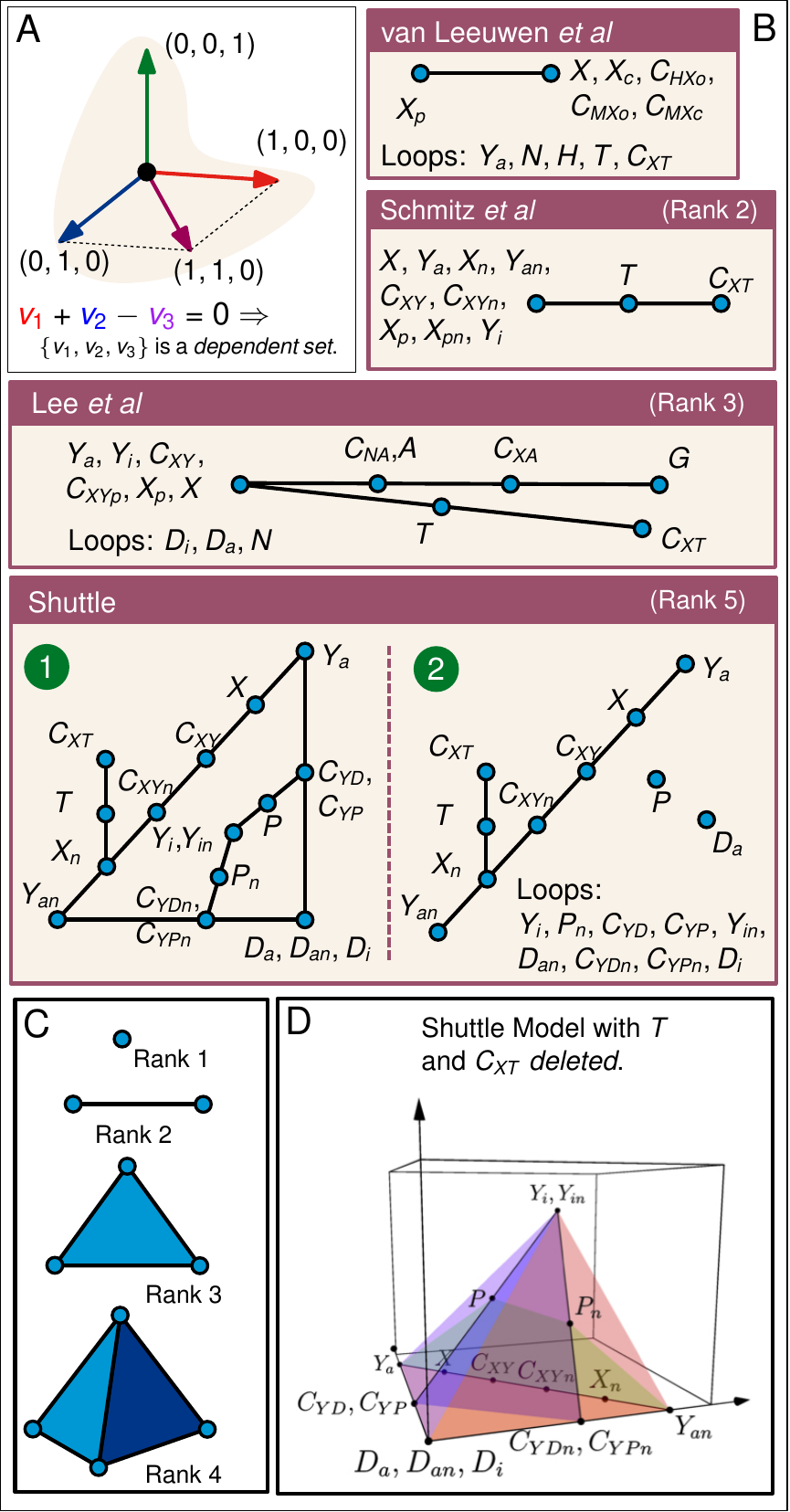}
\caption{Matroids allow for parameter-free model discrimination and prediction. {\bf(A)} Depiction of linear dependence and independence in a vector space (see main text for additional information). {\bf (B)} Schematic representation of the matroids for Wnt models. Each species represented by a loop is determined from the parameters alone; groups of species represented on a point can be determined by measurement of one of the species; groups of species represented on a line can be determined by measurement of two species. For notation used see Table S1.{\bf (C)} Schematic of rank, which corresponds to `what-to-measure'. So for rank 1, measure 1 species to determine all the others, for rank 2, measure 2, etc. {\bf(D)} 
Deletion is a matroid-theoretic operation which removes a species x from the ground set of the matroid and only considers dependencies of the original model that excluded x. Deleting $T$ and $C_{XT}$ gives a rank 4 matroid which can be visualized by planes in 3D space, as shown here.}
\label{fig-matroids}
\end{center}
\end{figure}

We calculate the matroid of five Wnt signaling pathway models (four shown in Figure \ref{fig-matroids}B). Each model has a \emph{rank} $r$, which dictates the number of species from the full set whose concentrations can be independently specified; taking measurements of $r$ independent species determines (in terms of parameters) the values for all other species. {\em Circuits} are minimal dependent sets of species -- they become important when we consider model discrimination. A matroid can be represented pictorially by point arrangements: the set of species labeling a point has rank $1$, the set on a line has rank $2$, the set on a $2$-dimensional plane has rank $3$, and so on (Figure \ref{fig-matroids}C). Any two species labeling the same point are algebraically related; as are any three species on a line, any four species in a plane, etc.
\par

We describe how the matroid of a model is computed in more detail in the SI; the input is the polynomial ideal of steady-state relations and the output is a list of all circuits with their polynomial relations. Strictly speaking, a circuit is defined as a set of variables. However, in this application, we record the polynomial relations, since these are the support-minimal steady-state invariants. One approach involves computing a Gr\"{o}bner basis for every elimination ordering, a feasible though lengthy computation for small systems. An alternative uses linear algebra to pinpoint the sets of variables appearing in invariants; then, it uses this information in conjunction with elimination or numerical algebraic geometry software to find polynomials (\cite{Rosen:2014} for more detail). This approach is only now being implemented because algebraic matroids have only recently been adopted for applications, e.g. low-rank matrix completion \cite{Kiraly:2012}.
\par
The results of the matroid calculations (Figure \ref{fig-matroids}B) prompt biological insight; for illustration, we analyze the van Leeuwen {\em et al.} model \cite{vanLeeuwen:2009ev}. In this model, five species (called {\em loops}) can be determined from just the parameters. Among the others, any pair not including $X_p$ (\bcat marked for ubiquitination) is dependent; therefore, an experiment measuring two of these concentrations could potentially reject the model if data are inconsistent with the relation. Assuming the model is consistent with data, measuring $X_p$ and any other non-loop is enough to determine all steady-state concentrations in terms of parameters.
\par
Unlike the other models, the solution set for the shuttle model has two irreducible components (loosely, proper subsets that should be considered separately). The matroids for the two irreducible components both have rank $5$. The number of components for a given model is determined by its algebraic structure; the number of components and real positive steady-states are not analogous  -- multiple steady-states may appear even when we have only one component. As described above, if we want to know all species concentrations in the shuttle model, $5$ measurements must be made and these measurements should be chosen to be independent. For example, measuring TCF ($T$) and any four species not lying in the same plane in Figure \ref{fig-matroids}D would determine all species concentrations. 
\par
The minimal sets of species from the matroid can also be used to study {\em part} of the system. The partial information (relations on a subset of species) obtained from the matroid can be used as a self-consistency check between data and competing models and, in this way, serves as a method to rule out models. Model discrimination based on steady-state invariants has been performed for one specific ordering, so by including the matroid we can recover all possible steady-state invariants with different variables (full set of circuit polynomials, as defined in \cite{Kiraly:2013}), including conserved quantities. For example, Tan {\em et al.} \cite{Tan:2014ii} measured average \bcat in the cytoplasm ($X$) and the nucleus ($X_n$) after Wnt stimulation, we can investigate the relationship between these species. There are two models that include $X$ and $X_n$ species: Schemitz {\em et al.} and shuttle. In the Schmitz {\em et al.} model, $X$ and $X_n$ is encoded in the circuit polynomial \[I = h_1(\delta) X^2 + h_2(\delta) X_n^2 + h_3(\delta) X X_n,\] where $h_i, i \in (1,2,3)$ are functions of the parameters ($\delta$). From the data in \cite{Tan:2014ii}, can test its compatibility with the model via a parameter-free method as described in \cite{Harrington:2012us}. Briefly, the method tests whether the data satisfy the Schmitz {\em et al.} circuit polynomial by checking whether there exist  $\tilde{h}_i$, $i \in (1,2,3)$ satisfying $I=0$, given these data.
Clearly, if the circuit polynomial is satisfied then the coplanarity condition holds with $\tilde{h}_i = h_i$.

Model compatibility is determined by computing the coplanarity error ($\Delta$) via the singular value decomposition of the matrix
\[ \left(
\begin{array}{ccc}
	& &  \\
	\hat{X}^2 & \hat{X}_n^2 & \hat{X}\hat{X}_n \\ 
	& & 
\end{array}
\right) \left(
\begin{array}{c}
	\tilde{h}_1 \\
	\tilde{h}_2 \\ 
	\tilde{h}_3 
\end{array}
\right) = 0, \]
where $\hat{X}$ denotes the observed value of species $X$. 
The null hypothesis that the model is compatible with the data can be rejected when the coplanarity error (normalized smallest singular value) is greater than a statistical bound as described in \cite{Harrington:2012us} and SI, which is determined by the Gaussian measurement noise in the data and the invariant structure.
\par
Before we test the models with data from \cite{Tan:2014ii}, we simulate data from both the shuttle and Schmitz {\em et al.} models. We draw random parameters from a lognormal distribution and then simulate 100 replicate measurements of ($X,X_n$) with noise (we perturb the data with noise $\sim 10^{-6}N(0,1)$). We test model compatibility at 5\% significance level; results of the coplanarity test identify that the Schmitz {\em et al.} model is incompatible with data generated by the shuttle model ($\Delta_{Schmitz} = 64820$, where the compatibility cutoff is $11.15$). Unsurprisingly, the Schmitz {\em et al.} model is compatible with data generated by itself ($\Delta_{Schmitz} = 3.768$, cutoff $=11.15$).  We use the three replicates of $X$ and $X_n$ from \cite{Tan:2014ii} at $t=120,240$ minutes, and assume the data are close to steady-state. Since the noise in the data are unknown, we test different noise levels and are able to rule out the Schmitz {\em et al.} model up to noise $\sim 10^{-4}N(0,1)$ (see SI). By consulting the matroid of the shuttle model, we find that $X$ and $X_n$ are independent, thus no circuit polynomial exists and any data are compatible with the shuttle model (an additional species is required to form a circuit polynomial). Thus as demonstrated, matroids guide experiment design to discriminate between models with minimal required measurements.

\section{Conclusions}
There is a wealth of mathematical and experimental research on Wnt signaling, aimed at understanding the pathway well enough to target Wnt-implicated diseases. There are two significant challenges to overcome. The disparity between models and data that we have highlighted via Bayesian inference prevents us from constraining parameter values
in a manner that often helps to elucidate mechanisms and predict function. The second challenge is the gap between {\em in vitro} and {\em in vivo} studies, and the corresponding differences in parameter values. This is supported by evidence on the variation in parameter estimates between na\"{i}ve and crowded (physiological) {\em in vitro} molecular experiments \cite{Aoki:2011ji, Aoki:2013jk}. To gain insight into these complex systems described by complicated models, we must evade this parameter problem.
\par
Parameter-free approaches can provide additional information about the $\beta$-catenin/Wnt pathway. Based on injectivity/multistationarity analysis, we find that the shuttle model predicts the possibility of a regulatory switch, acting early in the cell fate determination pathway. Other systems have also reported early checkpoints in cell fate signaling in activation of apoptosis through receptor-ligand binding \cite{Ho:2010hb,scott:n:2009}. 
We identify the possibility of important roles for spatial localization and degradation in cell fate switching. In the Erk pathway, it is also seen that either localization (via shuttling) or degradation by apoptosis is crucial for bistable switching, both mathematically and experimentally \cite{Li:2000fk, bagci:bj:2006, Harrington:2013gp, Michailovici:2014hv}. To our knowledge, we report for the first time that a combination of these processes governs the dynamical regime. 
\par
By computing the algebraic matroid of different Wnt models, we can characterize the dependencies between species. The matroid results enable us to guide experiments (which species to measure) to discriminate between models with data, all the while not requiring parameter values.  
The mechanisms for bistability identified above are of course not exclusive and one could imagine many other models exhibiting bistability via different mechanisms. This scenario could prompt new insight using our model discrimination framework, now with multiple bistable models.
\par
Given the current (and growing) complexity of models across a wide range of topics,  tools such as those  demonstrated here offer new means for testing models and for predictions to be made. In addition, we provide possible directions for future experimentation to narrow the gap between data and models, and, through our predictions, help to unravel the workings of the intricate and essential Wnt pathway.

\section{Materials and Methods}
\subsection{Bayesian inference}
Model selection in systems biology can be performed using Bayesian inference \cite{Kirk:2013hq}. Here we perform parameter inference for model selection using approximate Bayesian computation, which forgoes evaluation of the likelihood function and instead calculates the (here Euclidean) distance between model and data \cite{Liepe:2014iw}, implemented in the {\tt ABC Sys-Bio} package \cite{Liepe:2010eg}. For each model we compare the total free \bcat level (in some cases addition of two species) with the data provided by \cite{Tan:2014ii}.
\subsection{Injectivity}
Determining whether a model is capable of multiple responses can be tested using injectivity. The injectivity of each model was determined using CRNT toolbox \cite{CRNT-toolbox}; for those that were not injective (Schmitz {\em et al.} and shuttle), we computed the determinant of the Jacobian following \cite{Harrington:2013gp}, and analyzed the sign of the coefficients in Mathematica (Ver. 9.0; Wolfram Research, Champaign, IL). Bifurcation diagrams were computed using {\tt Oscill8} (Available at: oscill8.sourceforge.net/doc) and visualized with MATLAB (R2013a; The MathWorks, Natick, MA).

\subsection{Matroids}
Matroid computation is performed by structured variable elimination on eligible polynomial systems. This is carried out using symbolic algebra software {\tt Macaulay2} \cite{M2} with the aid of packages presented in \cite{Rosen:2014}. Code available at: http://math.berkeley.edu/~zhrosen/matroids.html.  When the set of steady-state solutions has multiple irreducible components, the matroid was computed for each in order to assess the independence structure in each regime. Isolated points in the solution set were not analyzed, as the matroid is trivial. Model discrimination was performed in Sage (Available at: https://cloud.sagemath.com) following the method presented in \cite{Harrington:2012us}.

\section{Acknowledgements}
We thank A Burgess and C Wee Tan for data and discussions about Wnt signaling. We also thank the anonymous reviewers as well as T Dale, E Feliu, A Fletcher, K Ho, PK Maini, E O'Neill, A Shiu, and B Sturmfels for helpful discussions and/or comments on the manuscript. HAH gratefully acknowledges funding from EPSRC Fellowship EP/K041096/1 and American Institute of Mathematics. ZR and HAH acknowledge funding from Royal Society International Exchanges Scheme 2014/R1 IE140219. ALM and HMB acknowledge funding from Human Frontiers in Science Program (RGP0039/2011). ALM, HMB and HAH acknowledge funding from King Abdullah University of Science and Technology (KAUST) KUK-C1-013-04.


\pagebreak

\pagebreak

\pagebreak

\pagebreak

\pagebreak

\begin{center}
{\LARGE Supplementary Materials for: \\[2mm]
Parameter-free methods distinguish Wnt pathway models and guide design of experiments}
\end{center}

\section*{Abbreviations}
APC, adenomatous polyposis coli; CRNT, chemical reaction network theory; DC, destruction complex; Dsh, Dishevelled; GSK, glycogen synthase kinase; ODE, ordinary differential equation; PP, protein phosphatase; TCF, T-cell factor


\section{A suite of models of canonical Wnt signaling capture different aspects of the pathway}
In this section we introduce the four existing mathematical models of Wnt signaling that we study throughout: using methods for Bayesian inference alongside techniques from chemical reaction network theory (CRNT), injectivity theory, and matroid theory. In doing so we characterize and compare the structure and behaviors of alternative models of the Wnt pathway.
\par
Each model can be represented by a set of chemical reactions in terms of the interacting variables (species), with the reactions characterized by rate constant parameters. These reactions are combined via ordinary differential equations (ODEs) that describe how each species changes over time. The effects of Wnt enter each model through parameters that control the relevant reactions, as we specify below.
\par
The models that we study focus on different aspects of Wnt signaling. For example, 
Lee {\em et al.}'s model \cite{Lee:2003bw}, focusses on the formation of the destruction complex (DC) from its constituent parts, and how its subsequent ability to degrade $\beta$-catenin is altered by the presence and absence of an external Wnt stimulus.  
van Leeuwen {\em et al.} \cite{vanLeeuwen:2007ga} simplify the dynamics of the DC (by neglecting some of its constituent parts) and instead focus on the dual roles of \bcat in cell-cell adhesion and the production of target genes. This model distinguishes two forms of \bcat (open and closed): both forms can bind to TCF to produce target genes but only the open form binds to adhesion molecules. The balance between open and closed forms is regulated by Wnt: high levels of Wnt lead to disaggregation of DC, as in \cite{Lee:2003bw}, and also increase levels of open $\beta$-catenin.  
Schmitz \etal \cite{Schmitz:2013kn} consider a single form of \bcat and a single form of DC (active): their model focuses on the effect that shuttling of \bcat and DC between the cytoplasm and the nucleus  has on TCF binding to \bcat in the nucleus.
Finally, based on an asymptotic analysis of Lee {\em et al.}'s model, Mirams {\em et al.} \cite{Mirams:2010ha} propose a single equation to describe the dynamics of \bcat. Their model accounts for the production and degradation of \bcat and assumes that Wnt regulates the rate at which \bcat binds to DC.

\par
In the subsections that follow, we state the ODEs that define the four models of interest. To facilitate their comparison, we introduce a standardised notation for the model variables (see Table  \ref{tab-notation}):
where possible, the same symbol denotes the same species across all models. 
\par 
In the subsequent section, we present a new model that combines elements of the models from literature with more recent knowledge about the pathway. Following this, we introduce and explain the methodologies that we use before applying them to the five models of interest.

\begin{table}[hb!]
\begin{center}
\begin{tabular}{|c|c|l|}
	\hline {\bf Symbol} & {\bf Species} & {\bf Forms} \\ \hline
	$X$ & \bcat 			& 	$X_o$ -- open form 	\\
						& & $X_c$ -- closed form  \\
						& & $X_{p}$ -- marked for proteasomal degradation  \\ \hline
	$Y$ & Destruction complex		& $Y_a$ -- active 	\\
		& (APC/Axin/GSK3$\beta$)	& $Y_i$ -- inactive 	\\ \hline
	$D$ & Dishevelled 		& $D_a$ -- active \\ 
						& & $D_i$ -- inactive 	\\ \hline
	$A$ & APC			&  	\\
	$N$ & Axin			&  	\\
	$G$ & GSK3$\beta$ 			& 	\\
	$T$ & TCF			& 	\\
	$H$ & Adhesion protein 	& 	\\
	$P$ & Phosphatase		& 	\\ 
	$E$ & Target gene		& 	\\ \hline
	$C$ 	& Complex		& $C_{XY}$ -- complex of X and Y (etc.)	\\ \hline
\end{tabular}
\caption{Definition of notation for the species used across models.}
\label{tab-notation}
\end{center}
\end{table}

\subsection{Lee \etal model}
In this model, now canonical for the Wnt signalling pathway, the dynamics of 15 species are described \cite{Lee:2003bw}. The model focuses on the assembly of the destruction complex from its constituent parts (APC, Axin and GSK3$\beta$) but does not distinguish between the nucleus and cytoplasm, assuming instead that all species are uniformly distributed throughout the cell. The reactions that define the model are

\begin{equation*}
\begin{aligned}[t]
D_i 			&\xrightarrow{ \alpha_1} D_a \\
D_a 			&\xrightarrow{ \alpha_2} D_i \\
Y_a 			&\xrightarrow{\alpha_3} Y_i \\
Y_i 			&\xrightarrow{\alpha_4} Y_a\\
D_a + Y_i	&\xrightarrow{\alpha_5} D_a + G + C_{NA} \\
G + C_{NA} 	&\xrightarrow{\alpha_6} Y_i \\
Y_i 			&\xrightarrow{\alpha_7} G + C_{NA} \\
A + N 		&\xrightarrow{\alpha_8} C_{NA} \\
C_{NA} 		&\xrightarrow{\alpha_9} A + N \\
Y_a + X 	&\xrightarrow{\alpha_{10}} C_{XY} \\
C_{XY}  	&\xrightarrow{\alpha_{11}} Y_a + X\\
C_{XY}		&\xrightarrow{\alpha_{12}} C_{XYp} \\
\end{aligned} \hspace{25mm}
\begin{aligned}[t]
C_{XYp}	&\xrightarrow{\alpha_{13}} X_p + Y_a \\
X_p			&\xrightarrow{\alpha_{14}} 		\emptyset \\
\emptyset	&\xrightarrow{\alpha_{15}} X \\
X			&\xrightarrow{\alpha_{16}}  \emptyset\\
\emptyset	&\xrightarrow{\alpha_{17}} N \\
N			&\xrightarrow{\alpha_{18}} \emptyset \\
X + T		&\xrightarrow{\alpha_{19}} C_{XT} \\
C_{XT}		&\xrightarrow{\alpha_{20}} X + T \\
X + A		&\xrightarrow{\alpha_{21}} C_{XA} \\
C_{XA}		&\xrightarrow{\alpha_{22}} X + A \\
\end{aligned}
\end{equation*}

\noindent and the corresponding ODEs, where, for example ${D_i}'$ denotes differentiation of $D_i$ with respect to time, are
\begin{align*}
	{D_i}' 	&= -\alpha_1 D_i + \alpha_2 D_a \\
	{D_a}' 	&=  \alpha_1D_i - \alpha_2D_a \\
	{Y_a}' 	&=  \alpha_3Y_i - \alpha_4Y_a - \alpha_{10}XY_a + \alpha_{11} C_{XY} + \alpha_{13} C_{XYp} \\
	{Y_i}' 	&= - \alpha_5D_aY_i - \alpha_3Y_i + \alpha_4Y_a + \alpha_6 G C_{NA} - \alpha_7 Y_i \\
	{G}' 	&= \alpha_5D_aY_i - \alpha_6 G C_{NA} + \alpha_7 Y_i \\
	{C_{NA}}' 	&= \alpha_5D_aY_i - \alpha_6 G C_{NA} + \alpha_7 Y_i + \alpha_8 N A - \alpha_9 C_{NA} \\
	{A}'	&= -\alpha_8 N A + \alpha_9 C_{NA} - \alpha_{21} X A + \alpha_{22} C_{XA} \\
	{C_{XY}}'	&= \alpha_{10} X Y_a - \alpha_{11} C_{XY} - \alpha_{12} C_{XY} \\
	{C_{XYp}}' 	&= \alpha_{12} C_{XY} - \alpha_{13} C_{XYp} \\
	{X_p}'	 	&= \alpha_{13} C_{XYp} - \alpha_{14} X_p \\
	{X}'		&= - \alpha_{10} XY_a  + \alpha_{11} C_{XY} + \alpha_{15} - \alpha_{16} X - \alpha_{19} X T + \alpha_{20} C_{XT} - \alpha_{21} X A + \alpha_{22} C_{XA} 	\\
	{N}'			&= - \alpha_8 N A + \alpha_9 C_{NA} + \alpha_{17} - \alpha_{18} N 	\\
	{T}'			&= - \alpha_{19} X T + \alpha_{20} C_{XT} 	\\
	{C_{XT}}'		&= \alpha_{19} X T - \alpha_{20} C_{XT}	\\
	{C_{XA}}'		&= \alpha_{21} X A - \alpha_{22} C_{XA}.
\end{align*}

\noindent The variable names are defined in Table \ref{tab-notation} and the parameters $\alpha_k$, $k \in (1,2,...,22)$ have been redefined from the original rate constants used in \cite{Lee:2003bw} so that they correspond to the reaction scheme shown above. Wnt enters the Lee {\em et al.} model through the parameter $\alpha_1$ that controls the activation of Dsh.

\clearpage

\subsection{van Leeuwen \etal model}
This model focuses on the competition between adhesive and transcriptional processes for \bcat  \cite{vanLeeuwen:2007ga}. The reactions on which the model is based are

\begin{equation*}
\begin{aligned}[t]
N			&\xrightarrow{\gamma_{1}}	Y_a 	\\
Y_a			&\xrightarrow{\gamma_{2}}	N 	\\
\emptyset   	&\xrightarrow{\gamma_{3}}	N	\\
N			&\xrightarrow{\gamma_{4}} 	\emptyset 	\\
X_i + Y_a 		&\xrightarrow{\gamma_{5}} 	X_p + Y_a		\quad (i=o,c)\\
X_p		 	&\xrightarrow{\gamma_{6}}	\emptyset	\\
\emptyset   	&\xrightarrow{\gamma_{7}}	X_o			\\
X_i   		&\xrightarrow{\gamma_{8}}	\emptyset	\quad (i=o,c) 	\\
X_o + H		&\xrightarrow{\gamma_{9}} 	C_{HX}	\\
C_{HX}	&\xrightarrow{\gamma_{10}} 	X_o + H		\\
\end{aligned} \hspace{25mm}
\begin{aligned}[t]
X_i + T		&\xrightarrow{\gamma_{11}} 	C_{XT_i}	 \quad (i=o,c) \\
C_{XT_i}	&\xrightarrow{\gamma_{12}} 	X_i + T	 \quad (i=o,c)	\\
X_o			&\xrightarrow{\gamma_{13}} 		X_c 		\\
\emptyset   	&\xrightarrow{\gamma_{14}}	H			\\
H	   		&\xrightarrow{\gamma_{15}}	\emptyset	\\
\emptyset   	&\xrightarrow{\gamma_{16}}	T			\\
T			&\xrightarrow{\gamma_{17}} 	\emptyset 	\\
C_{XT} 		&\xrightarrow{\gamma_{18}} 	E + C_{XT}	\\
E			&\xrightarrow{\gamma_{19}} 	\emptyset 	\\
Y_a			&\xrightarrow{\gamma_{20}} \emptyset		\\
\end{aligned}
\end{equation*}

\vspace{5mm} The 19 parameters $\gamma_k  \; (k=1,2,\ldots,19)$ relate to specific reactions. Note that four rate constants $(k=5,6,11,12)$ are common to both forms of \bcat --- van Leeuwen {\em et al.} assume that the rates of binding and phosphorylation of \bcat by the DC, and binding of \bcat to transcription molecules are equal for \bcat in its open ($X_o$) and closed ($X_c$) forms.  In the ODEs below we use the notation $X \equiv X_o$ so that the notation for free \bcat ($X$) is consistent across models. Thus the equations associated with these reactions are
\begin{align*}
	{Y_a}'		&= \gamma_1 N -  (\gamma_2 + \gamma_{20}) Y_a  \\
	{N}' 	   	&= - \gamma_1 N + \gamma_2 Y_a + \gamma_3 - \gamma_{4} N  \\
	{X_p}' 	&= \gamma_5 Y_a \frac{X}{X+X_c+\gamma_{5}^*} - \gamma_6 X_p +  \gamma_5 Y_a \frac{X_c}{X+X_c+\gamma_{5}^*}  \\
	{X}'   	&= \gamma_7 - \gamma_8 X - \gamma_5 Y_a \frac{X}{X+X_c+\gamma_{5}^*} - \gamma_9 X H + \gamma_{10} C_{HX} - \gamma_{11} X T + \gamma_{12} C_{XT} - \gamma_{13} \frac{X}{X + \gamma_{13}^*}  \\
	{X_c}'   	&= \gamma_{13} \frac{X}{X + \gamma_{13}^*} - \gamma_5 Y_a \frac{X_c}{X+X_c+\gamma_{5}^*} - \gamma_8 X_c - \gamma_{11} X_c T + \gamma_{12} C_{XTc}  \\
	{H}'     	&= \gamma_{14} - \gamma_{15} H - \gamma_9 X H + \gamma_{10} C_{HX}  \\
	{C_{HX}}' &= \gamma_9 X H - \gamma_{10} C_{HX}  \\ 
	{T}	'        &= - \gamma_{11} X T + \gamma_{12} C_{XT}  - \gamma_{11} X_c T + \gamma_{12} C_{XTc} + \gamma_{16} - \gamma_{17} T\\
	{C_{XT}}' &= \gamma_{11} X T - \gamma_{12} C_{XT}  \\
	{C_{XTc}}' &= \gamma_{11} X_c T - \gamma_{12} C_{XTc}  \\
	E	    &= \gamma_{18} \frac{(C_{XT}+C_{XTc})}{(C_{XT}+C_{XTc}+\gamma_{18}^*)} - \gamma_{19} E
\end{align*}
where the species are  given in Table \ref{tab-notation}. Wnt enters the model through its inhibition of the DC (by causing its dissociation and degradation), and its effects are encompassed by parameters $\gamma_2, \gamma_4$ and $\gamma_{20}$ above.

Species $E$ is called `target gene' by  van Leeuwen \etal Note that is is closely related it to the $\beta$-catenin-TCF complex that exists in this model and the other models under investigation ($C_{XT})$, since it is a downstream product of this complex that does not depend on any other species.

\subsection{Schmitz  \etal model}
The model of Schmitz {\em et al.} \cite{Schmitz:2013kn} considers interactions between \bcat and DC in the cytoplasm and nucleus. In each compartment, DC binding to \bcat leads to its phosphorylation, and phosphorylated \bcat is degraded. We use subscript {\em n} to denote species residing in the nucleus with the exception of TCF ($T$) and the $\beta$-catenin-TCF complex ($C_{XT}$): since these species are always localised in the nucleus (they do not shuttle), the subscript is omitted to facilitate comparison with the other models. The reactions that specify this model are

\begin{equation*}
\begin{aligned}[t]
\emptyset	&\xrightarrow{\delta_{0}}		X	\\
X      		&\xrightarrow{\delta_{1}} 	X_n	\\
X_n     		&\xrightarrow{\delta_{2}} 	X	\\
Y_a			&\xrightarrow{\delta_{3}} 	Y_{an}	\\
Y_{an}     		&\xrightarrow{\delta_{4}} 	Y_a	\\
X + Y_a        	&\xrightarrow{\delta_{5}} 	C_{XY}	\\
C_{XY} 		&\xrightarrow{\delta_{6}}	X + Y_a	\\
C_{XY} 		&\xrightarrow{\delta_{7}}	Y_a + X_p	\\
\end{aligned} \hspace{25mm}
\begin{aligned}[t]
X_n + Y_{an}  	&\xrightarrow{\delta_{8}} 	C_{XYn} 	\\
C_{XYn} 	&\xrightarrow{\delta_{9}}	X_n + Y_{an}	\\
C_{XYn} 	&\xrightarrow{\delta_{10}}	Y_{an} + X_{pn}	\\
X_n + T	  	&\xrightarrow{\delta_{11}} 	C_{XT} 		\\
C_{XT} 		&\xrightarrow{\delta_{12}}	X_n + T		\\
X_p 		&\xrightarrow{\delta_{13}} \emptyset	\\                          
X_{pn}		&\xrightarrow{\delta_{14}} \emptyset	\\                          
Y_a			&\xrightarrow{\delta_{15}} 	Y_i 		\\
Y_i			&\xrightarrow{\delta_{16}} 	Y_a 	\\
\end{aligned}
\end{equation*}
\par
\noindent and the associated ODEs are given by

\begin{align*}
	{X}'        	&=\delta_{0} - \delta_1 X + \delta_2 X_n - \delta_5 X Y_a + \delta_6 C_{XY}  \\
	{Y_a}'      	&=- \delta_3 Y_a + \delta_4 Y_{an} - \delta_5 X Y_a + (\delta_6+\delta_7) C_{XY} - \delta_{15} Y_a + \delta_{16} Y_i  \\
	{X_n}'	   	&= \delta_1 X - \delta_2 X_n - \delta_8 X_n Y_{an} + \delta_9 C_{XYn} - \delta_{11} X_n T + \delta_{12} C_{XT} \\
	{Y_{an}}'	 	&= \delta_3 Y_a - \delta_4 Y_{an} - \delta_8 X_n Y_{an} +  (\delta_9+\delta_{10}) C_{XYn}  \\
	{C_{XY}}'	&=\delta_5 X Y_a - (\delta_6+\delta_7) C_{XY}  \\
	{C_{XYn}}'	&=\delta_8 X_n Y_{an} - (\delta_9+\delta_{10}) C_{XYn}  \\
	{T}'	    	&= - \delta_{11} X_n T + \delta_{12} C_{XT}  \\
	{C_{XT}}'	&=\delta_{11} X_n T - \delta_{12} C_{XT}  \\
	{X_p}'	 	&= \delta_7 C_{XY} - \delta_{13} X_p  \\
	{X_{pn}}'	&=\delta_{10} C_{XYn} - \delta_{14} X_{pn}   \\
	{Y_i}'	 	&= \delta_{15} Y_a - \delta_{16} Y_i                                                   
\end{align*}
where $\delta_k \; (k=1,2,\ldots,17)$ are the non-negative constants associated with the reactions above. In this model Wnt acts to inactivate the destruction complex in the cytoplasm though parameter $\delta_{15}$.

\clearpage
\subsection{Mirams \etal model}
In \cite{Mirams:2010ha}, a mathematical analysis of Lee {\em et al.}'s model \cite{Lee:2003bw} is undertaken. This analysis reveals that the reactions occur on three, markedly different timescales: the shortest timescale  corresponds to phosphorylation of \bcat while bound to the DC; the intermediate timescale corresponds to activation/inactivation of the DC via a signal from Dsh; and the longest timescale corresponds to changes in levels of free \bcat (through degradation by the DC). By focussing on the longest timescale, the authors derive a reduced model, comprising a single differential equation, that is proposed to describe how the free \bcat level changes over time. In \cite{Mirams:2010ha} the authors study a time-dependent Wnt stimulus; here we do not include the Wnt time-dependence since it does not feature in any other models and we want to compare them as fairly as possible.  The reactions that are retained in this model are

\begin{equation*}
\begin{aligned}[t]
\emptyset 	&\xrightarrow{\zeta_{1}} 	X 		\\
X			&\xrightarrow{\zeta_{2}}	\emptyset
\end{aligned} \hspace{25mm}
\begin{aligned}[t]
X + T   		&\xrightarrow{\zeta_{3}}  	C_{XT} 	\\
C_{XT}  		&\xrightarrow{\zeta_{4}}  	X + T 	\\
X + N		&\xrightarrow{\zeta_{5}} 	\emptyset + N
\end{aligned}
\end{equation*}

\noindent where TCF ($T$) and Axin ($N$)  feature as interactants with \bcat ($X$), and are not modeled explicitly. Mirams  {\em et al.} retain the original parameterization of Lee {\em et al.} in their expression for \bcat whereas here we rename the parameters. The dynamics of $X$ over these timescales are thus governed by
\begin{align*}
	X' &= \zeta_1 - X \frac{\zeta_3'}{\zeta_4' (\zeta_5' + X)} - \zeta_2 X
\end{align*}
where $\zeta_k^{'} \; (k=3,4,5)$ are combinations of the rate constants $\zeta_i \; (i=3,4,5)$ and parameters from the Lee {\em et al.} model. The influence of Wnt, which in \cite{Mirams:2010ha} was time-dependent but now is assumed constant, is incorporated into the dynamics via parameter $\zeta_3'$.

\clearpage
\section{A new Wnt model focuses on the processes of shuttling and degradation}
 
We introduce a new model of the canonical Wnt signaling pathway in order to investigate recent biological findings in a theoretical context (see Figure 1A for a schematic diagram). The model distinguishes two cellular compartments: cytoplasm and nucleus. Species marked with subscript {\em n} reside in the nucleus and species without this subscript reside in the cytoplasm, with the exception of $T$ and $C_{XT}$, which remain localised in the nucleus and share their notation with the Schmitz \etal model (see section S1.3). 
\par
The central component of the shuttle model is the destruction complex (DC), here denoted by species $Y$. It is activated and inactivated by active Dishevelled ($D$) and Phosphatase ($P$), respectively. Active DC ($Y_a$) degrades \bcat in both cellular compartments, inhibiting its ability to bind to TCF ($T$) and promote downstream gene transcription via the $\beta$-catenin-TCF nuclear complex ($C_{XT}$). Inactive DC ($Y_i$) binds but does not degrade \bcat \cite{Li:2012jw}. The reactions that govern the shuttle model are given below and are summarized in Table \ref{tab-reactions}.

\begin{equation*}
\begin{aligned}[t]
 Y_a + X 		&\xrightarrow{k_{1}} 	C_{YX} 		\\
C_{YX} 			&\xrightarrow{k_{2}} Y_a + X			\\
C_{YX} 			&\xrightarrow{k_{3}} Y_a + \emptyset	\\
  \emptyset     	&\xrightarrow{k_{4}} 		X  			\\
X  				&\xrightarrow{k_{5}} \emptyset  \\
  Y_a + D_a 		&\xrightarrow{k_{6}} 	C_{YD} 	\\
  C_{YD} 		&\xrightarrow{k_{7}} Y_a + D_a	\\
C_{YD} 			&\xrightarrow{k_{8}} Y_i + D_a	\\			    
  Y_i + P  		&\xrightarrow{k_{9}} 	C_{YP} 		\\
  C_{YP} 		&\xrightarrow{k_{10}} Y_i + P		\\
  C_{YP} 		&\xrightarrow{k_{11}} Y_a + P		\\	    
Y_{an} + X_n 	&\xrightarrow{k_{12}} 	C_{YXn} \\
C_{YXn} 		&\xrightarrow{k_{13}} Y_{an} + X_n		\\
C_{YXn} 		&\xrightarrow{k_{14}} Y_{an} + \emptyset	
\end{aligned}	\hspace{25mm}
\begin{aligned}[t]
  X_n			&\xrightarrow{k_{15}}	\emptyset	\\
Y_{an} + D_{an}  	&\xrightarrow{k_{16}} 	C_{YDn}	\\
C_{YDn} 		&\xrightarrow{k_{17}} Y_{an} + D_{an} 	\\	
C_{YDn} 		&\xrightarrow{k_{18}} Y_{in} + D_{an} 	\\    
Y_{in} + P_n 	&\xrightarrow{k_{19}} 	C_{YPn} 	\\
C_{YPn} 		&\xrightarrow{k_{20}} Y_{in} + P_n 	\\
C_{YPn} 		&\xrightarrow{k_{21}} Y_{an} + P_n 	\\		    
   Y_i      		&\xrightarrow{k_{22}} 	Y_{in} 	\\
   Y_{in}   		&\xrightarrow{k_{23}} 	Y_i		\\
   X_n     			&\xrightarrow{k_{24}} 	X	  	\\
   X      			&\xrightarrow{k_{25}} 	X_n  	\\
   D_a                	&\xrightarrow{k_{26}}	D_{an}	\\		                
   D_{an}           	&\xrightarrow{k_{27}}	D_a		\\	
    D_i 			&\xrightarrow{k_{28}}	D_a		\\
 D_a 			&\xrightarrow{k_{29}}	D_i		\\
    X_n + T		&\xrightarrow{k_{30}}	C_{XT}	\\
C_{XT}			&\xrightarrow{k_{31}}	X_n + T		                
\end{aligned} 
\end{equation*}

\begin{table}[H]
\begin{center}
\begin{tabular}{|c|l|l|}
	\hline {\bf Reaction} & {\bf Explanation} & {\bf Evidence } \\ \hline
	 $D_i \rightleftarrows D_a$ 	& (in)activation of Dsh		& \cite{Lee:2003bw} \\ 
	$\emptyset \rightarrow X$  	& \bcat production			& \cite{Lee:2003bw, vanLeeuwen:2007ga, Mirams:2010ha} \\
	$X  \rightarrow \emptyset$	& DC-independent \bcat degradation	& \cite{Lee:2003bw, vanLeeuwen:2007ga, Mirams:2010ha} \\
	$Y_a + X 	\rightleftarrows 	C_{YX} 	\rightarrow Y_a + \emptyset$
	 						& DC-dependent \bcat degradation & \cite{Lee:2003bw, vanLeeuwen:2007ga, Schmitz:2013kn, Mirams:2010ha} \\
	$  Y_a + D_a 	\rightleftarrows 	C_{YD} 	\rightarrow Y_i + D_a$
							& DC open $\rightarrow$ closed & \cite{Li:2012jw} \\
	$Y_i + P \rightleftarrows C_{YP} 	\rightarrow Y_a + P$
							& DC closed $\rightarrow$ open & \cite{Li:2012jw} \\ \hline	    
	$X_n \rightarrow \emptyset$ & DC-independent \bcat degradation (nucleus)	&  \cite{Lee:2003bw, Schmitz:2013kn} 	\\ 
	$Y_{an} + X_n \rightleftarrows 	C_{YXn} 	\rightarrow Y_{an} + \emptyset$												& DC-dependent \bcat degradation (nucleus) & \cite{Schmitz:2013kn} \\
	$Y_{an} + D_{an}  \rightleftarrows C_{YDn} 	\rightarrow Y_{in} + D_{an}$
	    						& DC open $\rightarrow$ closed (nucleus) & \cite{Li:2012jw, Itoh:2005kg}\\
	$Y_{in} + P_n \rightleftarrows C_{YPn} \rightarrow Y_{an} + P_n$
							& DC closed $\rightarrow$ open (nucleus) & \cite{Li:2012jw, Itoh:2005kg}\\
	$X_n + T \rightleftarrows C_{XT}$ & \bcat binding to TCF (nucleus) & \cite{Lee:2003bw, vanLeeuwen:2007ga, Schmitz:2013kn, Mirams:2010ha, Schmitz:2011fe}  \\ \hline
	$Y_i \rightleftarrows Y_{in}$ & Shuttling of closed-form DC  & \cite{Schmitz:2013kn, Wiechens:2004fn, Cong:2004ew} \\
   	$X \rightleftarrows X_n$		& Shuttling of \bcat  & \cite{Schmitz:2013kn, Schmitz:2011fe}\\
   	$D_a \rightleftarrows D_{an}$ & Shuttling of active Dsh  & \cite{Itoh:2005kg} \\ \hline
\end{tabular}
\caption{Summary of the reactions included in the shuttle model. Evidence for each reaction is given via references from the literature: either a published model that has incorporated this or direct experimental evidence for the reaction.}
\label{tab-reactions}
\end{center}
\end{table}

The ODEs that correspond to these reactions are given by:
\begin{align*}
	{Y_a}'		&= - k_{1} Y_a X + (k_{2} + k_{3}) C_{XY} - k_{6} Y_a D_a + k_{7} C_{YD} + k_{11} C_{YP}  \\
	{X}' 	    	&= k_{4} - k_{5} X + k_{24} X_n - k_{25} X - k_{1} Y_a X + k_{2} C_{XY}  \\
	{C_{XY}}' 	&=  k_{1} Y_a X - (k_{2} + k_{3}) C_{XY}  \\
	{Y_i}' 		&= - k_{22} Y_i + k_{23} Y_{in} + k_{8} C_{YD} - k_{9} Y_i P + k_{10} C_{YP}  \\
	{D_a}' 		  &= - k_{26} D_a + k_{27} D_{an} + k_{28} D_i - k_{29} D_a - k_{6} Y_a D_a + (k_{7} + k_{8}) C_{YD}  \\
	{C_{YD}}' 	&=  k_{6} Y_a D_a - (k_{7} + k_{8}) C_{YD}  \\
	{P}'		     	&= - k_{9} Y_i P + (k_{10} + k_{11}) C_{YP}  \\
	{C_{YP}}' 	&= k_{9} Y_i P - (k_{10} + k_{11}) C_{YP}  \\
	{Y_{an}}'	&= - k_{12} Y_{an} X_n + (k_{13} + k_{14}) C_{XYn} - k_{16} Y_{an} D_{an} + k_{17} C_{YDn} + k_{21} C_{YPn}  \\
	{X_n}'		&= - k_{24} X_n + k_{25} X - k_{15} X_n - k_{12} Y_{an} X_n + k_{13} C_{XYn} - k_{30} X_n T + k_{31} C_{XT}  \\	
	{C_{XYn}}'	&= k_{12} Y_{an} X_n - (k_{13} + k_{14}) C_{XYn}  \\
	{Y_{in}}'	 &= k_{22} Y_i - k_{23} Y_{in} + k_{18} C_{YDn} - k_{19} Y_{in} P_n + k_{20} C_{YPn}  \\
	{D_{an}}'		 &= k_{26} D_a - k_{27} D_{an} - k_{16} Y_{an} D_{an} + (k_{17} + k_{18}) C_{YDn}  \\
	{C_{YDn}}' 	 &= k_{16} Y_{an} D_{an} - (k_{17} + k_{18}) C_{YDn}  \\
	{P_n}'		 &= - k_{19} Y_{in} P_n + (k_{20} + k_{21}) C_{YPn}   \\
	{C_{YPn}}'	 &=  k_{19} Y_{in} P_n - (k_{20} + k_{21}) C_{YPn}  \\
	{D_i}'		 &= - k_{28} D_i + k_{29} D_a  \\
	{T}'		 &= - k_{30} X_n T + k_{31} C_{XT}  \\
	{C_{XT}}'	 &= k_{30} X_n T - k_{31} C_{XT}
\end{align*}
where the species are defined as above in Table \ref{tab-notation} and the parameter set $k_i \; (i=1,2,\ldots,31)$ defines the rate constants of the model. Here Wnt acts through activation of Dsh (that, in turn, inhibits the degradation of \bcat by the DC), modeled by parameter $k_{28}$.


\clearpage
\section{Composite model}
In order to facilitate comparison between models and explain how parameters relate to their counterparts in other models, in this section we construct a composite model. In this model all of the reactions considered by each of the other models are combined; shown in Figure S1. By suitable reduction (See Tables \ref{tab-compos1} and \ref{tab-compos2}) each individual model is recovered. The composite is described by the set of 32 ODEs specified below.

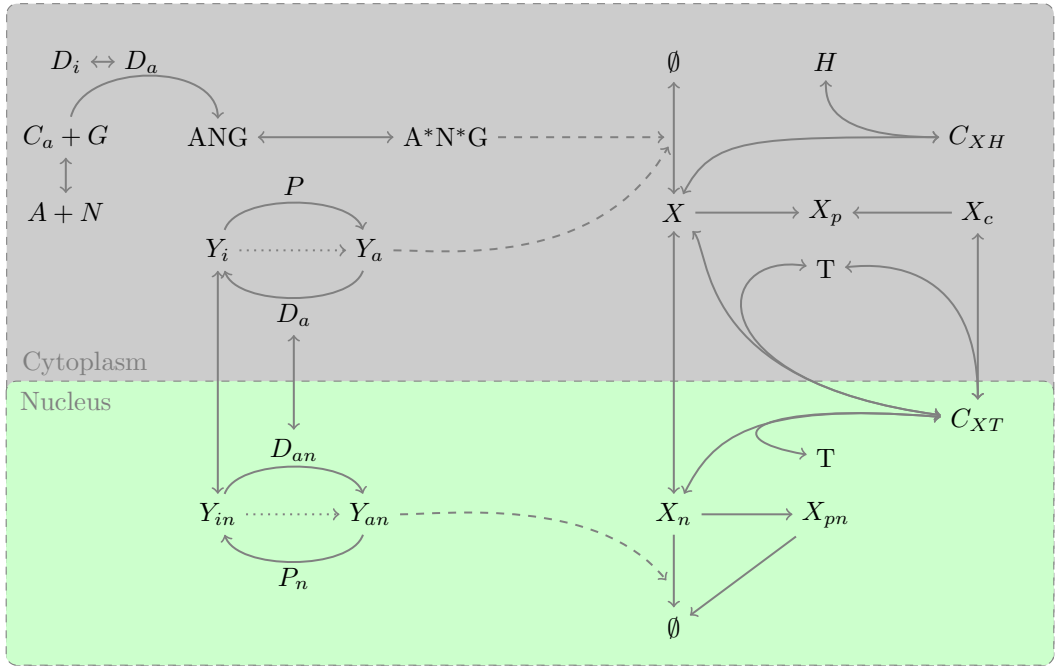
\begin{figure}[H]
\begin{center}
\begin{tikzpicture}

\node[rectangle,rounded corners] (AN) at (-4,10.5) {$A+N$};
\node[rectangle,rounded corners] (Ca) at (-4,11.5) {$C_a+G$};
\node[rectangle,rounded corners] (ANG) at (-2,11.5) {ANG};
\node[rectangle,rounded corners] (Dan) at (-1,7.35) {$D_{an}$};
\node[rectangle,rounded corners] (Yi) at (-2,10) {$Y_i$};
\node[rectangle,rounded corners] (Ya) at (0,10) {$Y_a$};
\node[rectangle,rounded corners] (Yin) at (-2,6.5) {$Y_{in}$};
\node[rectangle,rounded corners] (Yan) at (0,6.5) {$Y_{an}$};
\node[rectangle,rounded corners] (Cxt) at (8,7.75) {$C_{XT}$};
\node[rectangle,rounded corners] (empt) at (4,12.5) {$\emptyset$};
\node[rectangle,rounded corners] (An) at (1,11.5) {A$^*$N$^*$G};
\node[rectangle,rounded corners]  (T) at (6,9.75) {T};
\node[rectangle,rounded corners]  (Tn) at (6,7.25) {T};
\node[rectangle,rounded corners] (X) at (4,10.5) {$X$};
\node[rectangle,rounded corners] (Xp) at (6,10.5) {$X_p$};
\node[rectangle,rounded corners] (Xc) at (8,10.5) {$X_c$};
\node[rectangle,rounded corners] (Xn) at (4,6.5) {$X_n$};
\node[rectangle,rounded corners] (Xpn) at (6,6.5) {$X_{pn}$};
\node[rectangle,rounded corners] (Cxh) at (8,11.5) {$C_{XH}$};
\node[rectangle,rounded corners] (H) at (6,12.5) {$H$};
\node[rectangle,rounded corners] (P) at (-1,10.85) {$P$};
\node[rectangle,rounded corners] (Pn) at (-1,5.65) {$P_n$};
\node[rectangle,rounded corners] (emptn) at (4,5) {$\emptyset$};
\node[rectangle,rounded corners] (Dshi) at (-4,12.5) {$D_i$};
\node[rectangle,rounded corners] (Dsha) at (-3,12.5) {$D_a$};
\node[rectangle,rounded corners] (Da2) at (-1,9.15) {$D_{a}$};
\node[rectangle,rounded corners] (D0) at (4,11.5) {};
\node[rectangle,rounded corners] (D00) at (4,5.5) {};
\draw[<->,draw=black!50,line width=.75pt] (X) -- (empt);
\draw[->,draw=black!50,line width=.75pt,dashed] (An) -- (D0);
\draw[->,draw=black!50,line width=.75pt] (X) -- (Xp);
\draw[->,draw=black!50,line width=.75pt] (Xn) -- (Xpn);
\draw[<->,draw=black!50,line width=.75pt] (Yi) -- (Yin);
\draw[<->,draw=black!50,line width=.75pt] (X) -- (Xn);
\draw[<->,draw=black!50,line width=.75pt] (Da2) -- (Dan);
\draw[<-,draw=black!50,line width=.75pt] (Xp) -- (Xc);
\draw[->,draw=black!50,line width=.75pt] (Xn) -- (emptn);
\draw[->,draw=black!50,line width=.75pt] (Xpn) -- (emptn);
\draw[<->,draw=black!50,line width=.75pt] (Ca) -- (AN);
\draw[<->,draw=black!50,line width=.75pt] (ANG) -- (An);
\draw[<->,draw=black!50,line width=.75pt] (Dshi) -- (Dsha);
\draw[->,draw=black!50,line width=.75pt,dotted] (Yi) -- (Ya);
\draw[->,draw=black!50,line width=.75pt,dotted] (Yin) -- (Yan);
\draw[<->,draw=black!50,line width=.75pt] (Cxt) -- (Xc);
\draw[ ->,draw=black!50,line width=.75pt] (Ca) .. controls(-3.75,12.5) and (-2, 12.5)..(ANG);
\draw[ <->,draw=black!50,line width=.75pt] (Cxt) .. controls(8,10) and (6.5,9.8)..(T);
\draw[ <->,draw=black!50,line width=.75pt] (X) .. controls(4.5,11.5) and (5, 11.5)..(Cxh);
\draw[ <->,draw=black!50,line width=.75pt] (H) .. controls(6,12) and (6, 11.5)..(Cxh);
\draw[ ->,draw=black!50,line width=.75pt] (Yi) .. controls(-1.75,10.75) and (-.25, 10.75)..(Ya);
\draw[ <-,draw=black!50,line width=.75pt] (Yi) .. controls(-1.75,9.25) and (-.25, 9.25)..(Ya);
\draw[ ->,draw=black!50,line width=.75pt] (Yin) .. controls(-1.75,7.25) and (-.25, 7.25)..(Yan);
\draw[ <-,draw=black!50,line width=.75pt] (Yin) .. controls(-1.75,5.75) and (-.25, 5.75)..(Yan);
\draw[ ->,draw=black!50,line width=.75pt,dashed] (Ya) .. controls(1,10) and (3, 9.75)..(D0);
\draw[ ->,draw=black!50,line width=.75pt,dashed] (Yan) .. controls(1,6.5) and (3, 6.75)..(D00);
\draw[ <->,draw=black!50,line width=.75pt] (X) .. controls(4.5,10) and (4.25, 8.25)..(Cxt);
\draw[ <->,draw=black!50,line width=.75pt] (Xn) .. controls(4.5,7.5) and (5, 8)..(Cxt);
\draw[ <->,draw=black!50,line width=.75pt] (T) .. controls(4.5,10) and (4.25, 8.25)..(Cxt);
\draw[ <->,draw=black!50,line width=.75pt] (Tn) .. controls(4.5,7.5) and (5, 8)..(Cxt);
\node[draw=none,fill=none,text=gray](cyt) at (-3.75,8.5) {Cytoplasm};
\node[draw=none,fill=none,text=gray](nuc) at (-4,8) {Nucleus};
\begin{pgfonlayer}{background}
       \draw[fill=black!20,rounded corners, draw=black!50, dashed]
            (Dshi.west |- Cxh.north)+(-0.45,1.5) rectangle (+9,4.5);     
        \draw[fill=green!20,rounded corners, draw=black!50, dashed]
            (Yi.west |- emptn.north)+(-2.5,3) rectangle (+9,4.5);           
    \end{pgfonlayer}
\end{tikzpicture}
\end{center}
	\label{fig-composite}
	\caption{Reaction scheme of the composite model, which includes all species and reactions that are described in the previous models. Solid arrows denote direct reactions; long-dashed arrows denote species that act as catalysts in degradation reactions; and dotted arrows denote alternative paths for the direct activation of $Y$. Note that active/inactive forms of $Y$ are equivalent to active/inactive forms of ANG. }
\end{figure}

\begin{align*}
	{Y_a}'		&= - \lambda_{1} Y_a X + \lambda_{2} C_{XY} - \lambda_{3} Y_a D_a + \lambda_{4} C_{YD} + \lambda_{5} C_{YP} + \lambda_6 Y_i - \lambda_7 Y_a + \lambda_{8} C_{XYp} + \lambda_{9} N + \lambda_{10} Y_{an}   \\
	{Y_i}' 		&= - \lambda_{11} Y_i + \lambda_{12} Y_{in} + \lambda_{13} C_{YD} - \lambda_{14} Y_i P + \lambda_{15} C_{YP}  -\lambda_{16} D_aY_i - \lambda_{17} Y_i + \lambda_{18} Y_a + \lambda_{20} G C_{NA} \\
	{Y_{an}}'	&= - \lambda_{21} Y_{an} X_n + \lambda_{22} C_{XYn} - \lambda_{23} Y_{an} D_{an} + \lambda_{24} C_{YDn} + \lambda_{25} C_{YPn} + \lambda_{26} Y_a - \lambda_{10} Y_{an} \\
	{Y_{in}}'	 &= \lambda_{11} Y_i - \lambda_{12} Y_{in} + \lambda_{27} C_{YDn} - \lambda_{28} Y_{in} P_n + \lambda_{29} C_{YPn}  \\
	{X}' 	    	&= \lambda_{30} - \lambda_{31} X - \lambda_{32} X + \lambda_{33} X_n -  \lambda_{34}Y_a X - \lambda_{35}\frac{ Y_a X}{X + X_c + \lambda_{36}} + \lambda_{37} C_{XY}  - \lambda_{38} X T + \lambda_{39} C_{XT} \\&- \lambda_{40} X A + \lambda_{41} C_{XA} - \lambda_{42} X H + \lambda_{43} C_{HX} - \lambda_{44} \frac{X}{X + \lambda_{45}} - \lambda_{46} \frac{X}{\lambda_{47}(\lambda_{48} + X)}  \\
	{X_n}'		&= \lambda_{32} X - \lambda_{33} X_n - \lambda_{49} X_n - \lambda_{50} Y_{an} X_n + \lambda_{51} C_{XYn} - \lambda_{52} X_n T + \lambda_{53} C_{XT}  \\
	{X_c}'   	&=  \lambda_{44} \frac{X}{X + \lambda_{45}} -  \lambda_{54}\frac{ Y_a X_c}{X+X_c+ \lambda_{55}} - \lambda_{56} X_c - \lambda_{57} X_c T + \lambda_{58} C_{TX_c}  \\
	{X_p}'	 	&= \lambda_{8} C_{XYp} + \lambda_{35}\frac{Y_a X}{X+X_c+\lambda_{36}} +  \lambda_{35} \frac{Y_a X_c}{X+X_c+\lambda_{36}} + \lambda_{60} C_{XY}  - \lambda_{59} X_p \\
	{X_{pn}}'	&=\lambda_{61} C_{XYn} - \lambda_{62} X_{pn}   \\
	{G}' 	&=  \lambda_{16} D_aY_i - \lambda_{20}G C_{NA} + \lambda_{63}Y_i \\
	{C_{NA}}' 	&=  \lambda_{16} D_aY_i -  \lambda_{20} G C_{NA} +  \lambda_{63} Y_i +  \lambda_{64} N A -  \lambda_{65} C_{NA} \\
	{A}'	&= -\lambda_{64} N A + \lambda_{65}  C_{NA} - \lambda_{40} X A + \lambda_{41}C_{XA} \\
	{N}'			&= -\lambda_{64} N A + \lambda_{65}C_{NA} +\lambda_{66} - \lambda_{67}N + \lambda_{19}Y_a - \lambda_{9}N 	\\
	{H}'     	&= \lambda_{68} - \lambda_{69} H -  \lambda_{42} X H +  \lambda_{43}  C_{HX}  \\
	{D_a}' 		  &= - \lambda_{70} D_a + \lambda_{71} D_{an} + \lambda_{72} D_i - \lambda_{73} D_a - \lambda_{3} Y_a D_a + (\lambda_{4} + \lambda_{13}) C_{YD}  \\
	{D_i}'		 &= - \lambda_{72} D_i + \lambda_{73} D_a \\
	{D_{an}}'		 &= \lambda_{70} D_a - \lambda_{71} D_{an} - \lambda_{23} Y_{an} D_{an} + (\lambda_{24} + \lambda_{27}) C_{YDn}  \\
	{P}'		     	&= - \lambda_{14} Y_i P + (\lambda_{15} + \lambda_{5}) C_{YP}  \\
	{P_n}'		 &= - \lambda_{28} Y_{in} P_n + (\lambda_{29} + \lambda_{25}) C_{YPn}   \\
	{T}'		 	&= - \lambda_{38} X T + \lambda_{39} C_{XT} - \lambda_{52} X_n T + \lambda_{53} C_{XT}  - \lambda_{57} X_c T + \lambda_{58} C_{TX_c} + \lambda_{74} - \lambda_{75} T  \\
	{E}'			&= \lambda_{78} \frac{(C_{XT}+C_{XTc})}{(C_{XT}+C_{XTc}+\lambda_{79})} - \lambda_{80} E \\
	{C_{XT}}'	 &=  \lambda_{38} X T - \lambda_{39} C_{XT} + \lambda_{52} X_n T - \lambda_{53} C_{XT}  \\
	{C_{XTc}}'	&= \lambda_{57} X_c T - \lambda_{58} C_{TX_c}	\\ 
	{C_{XY}}' 	&=  \lambda_{1} Y_a X - \lambda_{2}C_{XY}  - \lambda_{76} C_{XY} \\
	{C_{XYn}}'	&= \lambda_{21} Y_{an} X_n - \lambda_{22} C_{XYn} \\
	{C_{XYp}}' 	&= \lambda_{76} C_{XY} - \lambda_{77} C_{XYp} \\
	{C_{YD}}' 	&=  \lambda_{3} Y_a D_a - (\lambda_{4} + \lambda_{13}) C_{YD}  \\
	{C_{YDn}}' 	 &= \lambda_{23} Y_{an} D_{an} - (\lambda_{24} + \lambda_{27}) C_{YDn}  \\
	{C_{YP}}' 	&= \lambda_{14} Y_i P - (\lambda_{15} + \lambda_{5}) C_{YP}  \\
	{C_{YPn}}'	 &=  \lambda_{28} Y_{in} P_n - (\lambda_{29} + \lambda_{25}) C_{YPn}  \\
	{C_{XA}}'		&= \lambda_{40} X A - \lambda_{41} C_{XA}	\\
 	{C_{HX}}' &= \lambda_{42} X H - \lambda_{43} C_{HX}  \\ 
\end{align*}
where $\lambda_i  \; (i=1,2,\ldots,80)$ parameterise the composite model; in Tables \ref{tab-compos1} and \ref{tab-compos2} the correspondence between $\lambda_i$ and the parameters of each of the other models is given.

\begin{table}
\begin{center}
\begin{tabular}{|c|c|c|c|c|c|}\hline
	Composite 		& Lee 	& van Leeuwen 	& Schmitz 	& Mirams 	&  Shuttle   \\\hline
	$\lambda_{1}$ 	& $\alpha_{10}$	& 0		& $\delta_5$	& 0			& $k_1$  	\\
	$\lambda_{2}$ 	& $\alpha_{11}$ 	& 0		& $\delta_6+\delta_7$	& 0			& $k_2 + k_3$ 	\\
	$\lambda_{3}$		& 0		& 0 		& 0			& 0 			& $k_6$	\\ 
	$\lambda_{4}$		& 0		& 0 		& 0 			& 0 			& $k_7$	\\ 
	$\lambda_{5}$		& 0		& 0 		& 0 			& 0 			& $k_{11}$		\\ 
	$\lambda_{6}$		& $\alpha_3$	& 0 		& $\delta_{16}$ 			& 0 			& 0 		\\ 
	$\lambda_{7}$		& $\alpha_4$	& $\gamma_2+\gamma_{20}$ 	& $\delta_3+\delta_{15}$ 	& 0 			& 0 		\\ 
	$\lambda_{8}$		& $\alpha_{13}$	& 0 		& 0 			& 0 			& 0 		\\ 
	$\lambda_{9}$	& 0		& $\gamma_1$ 		& 0 			& 0 			& 0 		\\ 
	$\lambda_{10}$	& 0		& 0 		& $\delta_4$	& 0 			& 0 		\\ 
	$\lambda_{11}$	& 0		& 0 		& 0 			& 0 			& $k_{22}$	\\ 
	$\lambda_{12}$	& 0		& 0 		& 0 			& 0 			& $k_{23}$	\\ 
	$\lambda_{w13}$	& 0		& 0 		& 0 			& 0 			& $k_{8}$	\\ 
	$\lambda_{14}$	& 0		& 0 		& 0 			& 0 			& $k_{9}$	\\ 
	$\lambda_{15}$	& 0		& 0 		& 0 			& 0 			& $k_{10}$	\\ 
	$\lambda_{16}$	& $\alpha_5$	& 0 		& 0 			& 0 			& 0		\\ 
	$\lambda_{17}$	& $\alpha_3+\alpha_7$	& 0 	& $\delta_{16}$ 		& 0 			& 0		\\ 
	$\lambda_{18}$	& $\alpha_4$	& 0		& $\delta_{15}$	& 0 			& 0		\\ 
	$\lambda_{19}$	& 0		& $\gamma_2$ 		& 0	& 0 			& 0		\\ 
	$\lambda_{20}$	& $\alpha_6$	& 0 		& 0 			& 0 			& 0		\\ 
	$\lambda_{21}$	& 0		& 0 		& $\delta_8$		& 0 		& $k_{12}$		\\ 
	$\lambda_{22}$	& 0		& 0 		& $\delta_9+\delta_{10}$	& 0 		& $k_{13}+k_{14}$		\\ 
	$\lambda_{23}$	& 0		& 0 		& 0		& 0 		& $k_{16}$		\\ 
	$\lambda_{24}$	& 0		& 0 		& 0		& 0 		& $k_{17}$		\\ 
	$\lambda_{25}$	& 0		& 0 		& 0		& 0 		& $k_{21}$		\\ 
	$\lambda_{26}$	& 0		& 0 		& $\delta_3$		& 0 		& 0		\\ 
	$\lambda_{27}$	& 0		& 0 		& 0		& 0 		& $k_{18}$		\\ 
	$\lambda_{28}$	& 0		& 0 		& 0		& 0 		& $k_{19}$		\\ 
	$\lambda_{29}$	& 0		& 0 		& 0		& 0 		& $k_{20}$		\\ 
	$\lambda_{30}$	& $\alpha_{15}$	& $\gamma_7$ & $\delta_0$	& $\zeta_1$ 		& $k_{4}$		\\ 
	$\lambda_{31}$	& $\alpha_{16}$	& $\gamma_8$ & 0		& $\zeta_2$ 	& $k_{5}$		\\ 
	$\lambda_{32}$	& 0		& 0 		& $\delta_1$		& 0 		& $k_{25}$		\\ 	
	$\lambda_{33}$	& 0		& 0 		& $\delta_2$		& 0 		& $k_{24}$		\\ 
	$\lambda_{34}$	& $\alpha_{10}$	& 0 		&	$\delta_5$		& 0 		& $k_{1}$		\\ 
	$\lambda_{35}$	& 0	& $\gamma_5$	& 0		& 0 		& 0		\\ 
	$\lambda_{36}$	& 0	& $\gamma_5^*$	& 0		& 0 		& 0		\\ 
	$\lambda_{37}$	& $\alpha_{11}$	& 0	&	$\delta_6$		& 0 		& $k_{2}$		\\ 
	$\lambda_{38}$	& $\alpha_{19}$	& $\gamma_{11}$	&	0	& 0 		& 0		\\ 
	$\lambda_{39}$	& $\alpha_{20}$	& $\gamma_{12}$	&	0	& 0 		& 0		\\ 
	$\lambda_{40}$	& $\alpha_{21}$	& 0		&	0	& 0 		& 0		\\ \hline
\end{tabular}
\caption{Parameter conversion from composite model to other models part 1. (See Table \ref{tab-compos2}.)}
\label{tab-compos1}
\end{center}
\end{table}

\begin{table}
\begin{center}
\begin{tabular}{|c|c|c|c|c|c|}\hline
	Composite 		& Lee 	& van L 	& Schmitz 	& Mirams 	&  Shuttle   \\\hline
	$\lambda_{41}$	& $\alpha_{22}$	& 0		&	0	& 0 		& 0		\\ 
	$\lambda_{42}$	& 0	& $\gamma_{9}$		&	0	& 0 		& 0		\\ 
	$\lambda_{43}$	& 0	& $\gamma_{10}$		&	0	& 0 		& 0		\\ 
	$\lambda_{44}$	& 0	& $\gamma_{13}$		&	0	& 0 		& 0		\\ 
	$\lambda_{45}$	& 0	& $\gamma_{13}^*$		&	0	& 0 		& 0		\\ 
	$\lambda_{46}$	& 0	& 0		&	0	& $\zeta_{3}^{'}$ 		& 0		\\ 
	$\lambda_{47}$	& 0	& 0		&	0	& $\zeta_{4}^{'}$ 		& 0		\\ 
	$\lambda_{48}$	& 0	& 0		& 0	& $\zeta_{5}^{'}$ 		& 0		\\ 
	$\lambda_{49}$	& 0	& 0		&	0	& 0		& $k_{15}$		\\ 
	$\lambda_{50}$	& 0	& 0		& $\delta_8$	& 0		& $k_{12}$		\\ 
	$\lambda_{51}$	& 0	& 0		& $\delta_9$	& 0		& $k_{13}$		\\ 
	$\lambda_{52}$	& 0	& 0		& $\delta_{11}$	& 0		& $k_{30}$	\\ 
	$\lambda_{53}$	& 0	& 0		& $\delta_{12}$	& 0		& $k_{31}$	\\
	$\lambda_{54}$	& 0	& $\gamma_{5}$	& 0	& 0		& 0		\\ 
	$\lambda_{55}$	& 0	& $\gamma_{5}^*$	& 0	& 0		& 0		\\ 
	$\lambda_{56}$	& 0	& $\gamma_{8}$	& 0	& 0		& 0		\\ 
	$\lambda_{57}$	& 0	& $\gamma_{11}$	& 0	& 0		& 0		\\ 
	$\lambda_{58}$	& 0	& $\gamma_{12}$	& 0	& 0		& 0		\\ 
	$\lambda_{59}$	& $\alpha_{14}$	& $\gamma_6$	& $\delta_{13}$	& 0		& 0		\\ 
	$\lambda_{60}$	& 0	& 0	& $\delta_{7}$	& 0		& 0		\\ 
	$\lambda_{61}$	& 0	& 0	& $\delta_{10}$	& 0		& 0		\\ 
	$\lambda_{62}$	& 0	& 0	& $\delta_{14}$	& 0		& 0		\\ 
	$\lambda_{63}$	& $\alpha_7$	& 0	& 0	& 0		& 0		\\ 
	$\lambda_{64}$	& $\alpha_8$	& 0	& 0	& 0		& 0		\\ 
	$\lambda_{65}$	& $\alpha_9$	& 0	& 0	& 0		& 0		\\ 
	$\lambda_{66}$	& $\alpha_{17}$	& $\gamma_{3}$	& 0	& 0		& 0		\\ 
	$\lambda_{67}$	& $\alpha_{18}$	& $\gamma_{4}$	& 0	& 0		& 0		\\ 
	$\lambda_{68}$	& 0	& $\gamma_{14}$	& 0	& 0		& 0		\\ 
	$\lambda_{69}$	& 0	& $\gamma_{15}$	& 0	& 0		& 0		\\ 
	$\lambda_{70}$	& 0	& 0	& 0	& 0		& $k_{26}$		\\ 
	$\lambda_{71}$	& 0	& 0	& 0	& 0		& $k_{27}$		\\ 
	$\lambda_{72}$	& $\alpha_1$	& 0	& 0	& 0		& $k_{28}$		\\ 
	$\lambda_{73}$	& $\alpha_2$	& 0	& 0	& 0		& $k_{29}$		\\ 
	$\lambda_{74}$	& 0	& $\gamma_{16}$	& 0	& 0		& 0		\\ 
	$\lambda_{75}$	& 0	& $\gamma_{17}$	& 0	& 0		& 0		\\ 
	$\lambda_{76}$	& $\alpha_{12}$	& 0	& 0	& 0		& 0		\\ 
	$\lambda_{77}$	& $\alpha_{13}$	& 0	& 0	& 0		& 0		\\ 
	$\lambda_{78}$	& 0	& $\gamma_{18}$	& 0	& 0		& 0		\\ 
	$\lambda_{79}$	& 0	& $\gamma_{18}^*$	& 0	& 0		& 0		\\ 
	$\lambda_{80}$	& 0	& $\gamma_{19}$	& 0	& 0		& 0		\\ \hline
\end{tabular}
\caption{Parameter conversion from composite model to other models part 2. (See Table \ref{tab-compos1}.)}
\label{tab-compos2}
\end{center}
\end{table}

\clearpage

\section{Extended methods and analysis I: Bayesian inference}
\subsection{Approximate Bayesian computation with {\tt ABC-SysBio}}

As described in the main text, we used approximate Bayesian computation (ABC) to infer a subset of parameter values for each model. Parameter inference is an important test of any model; we want to see whether the model provides a good fit to the data and, if so, to identify parameter ranges that can give rise to realistic fits. The key advantages of Bayesian inference over methods that give point estimates for the parameters or frequentist methods are:
\begin{itemize}
\item providing a distribution over parameter space that describes those parameter regions that have a high probability of having generated the data (the posterior distribution), and
\item taking into account prior knowledge about the system.
\end{itemize} 
As such it allows the modeler to hone in on regions of parameter space that are of interest, and ignore those that are not. Furthermore, the posterior distribution gives information about joint distributions in parameter space and can reveal multivariate dependencies between parameters.
\par
The posterior distribution can be described starting from Bayes rule:
\begin{align*} 
P(\theta|x) &\propto P(x|\theta)P(\theta)
\end{align*}
where $\theta$ represents a parameter set that describes the model and $x$ represents data with which we will compare the model. $P(\theta|x)$, the probability of $\theta$ given $x$, is called the posterior probability, $P(x|\theta)$ is the likelihood function and $P(\theta)$ is the prior probability (that is, knowledge we have about parameters before we begin) \cite{Gelman14}. As well as the full (joint) posterior distribution, one may also analyse the marginal posterior distributions which are the individual distributions over each parameter.
\par
In ABC, we forego evaluation of the likelihood function and instead compare the real data ($x$) with data simulated from the underlying probability model, denoted $x_m$. If the underlying model is given by $f = f(x_m|\theta)$, then we express the ABC posterior function by
\begin{align*} 
P_{ABC}(\theta|x) &\propto  \mathds{1}(\Delta(x,x_m)\le \epsilon) f(x_m|\theta) P(\theta)
\end{align*}
where $\Delta(a,b)$ denotes a distance measure between $a$ and $b$ and $\epsilon$ is the tolerance level that determines how well real and simulated data should agree \cite{Toni:2009gm}.
\par
ABC for parameter inference has been implemented in the software package {\tt ABC-SysBio} with support for parallelization \cite{Liepe:2010eg}. For the parameter inference performed in this work, we used the CUDA implementation of {\tt ABC-SysBio} with a Euclidean distance measure between model and data \cite{Liepe:2014iw}.  
\par
The models of Wnt signaling studied here differ in size and complexity, ranging from 1 (Mirams) to 19 (shuttle) species. In order to compare these models via parameter inference with ABC, we chose to limit the number of parameters that could vary to 3 for each model. We fixed all other parameters to be constant, at values that were previously used with these models (either estimated or calculated from experiment). With the exception of Lee {\em et al.}'s model, where half of the parameters were measured, the majority of parameters for the other published models were estimated from theory or from other models.  Limiting the number of free parameters in this way has two important consequences. First, it facilitates their comparison, since the dimension of the parameter space is the same in each case, and second, it reduces the number of behaviors each model can perform. If a reduced version of a model can fit the data well, as is the case here, performing model selection on more detailed models will not be possible and model comparison via parameter inference may yield little insight.

\subsection{Results and discussion of ABC parameter inference}
Using the composite model introduced above, we have selected parameters to infer for each of the five  models under analysis by their impact on \bcat signaling dynamics. 
Parameter inference using ABC outputs the posterior distribution over the model parameters. One can then simulate the model using parameters sampled from this posterior distribution and assess the fit between simulated trajectory and the data. These fits are shown in the main text in Figure 3C. Here we discuss the posterior distributions that gave rise to these fits. The marginal posterior distributions for each of the three parameters fit per model are shown in Figure \ref{fig-posts}.

\begin{figure}[H]
\begin{center}
\includegraphics[width=0.5\textwidth]{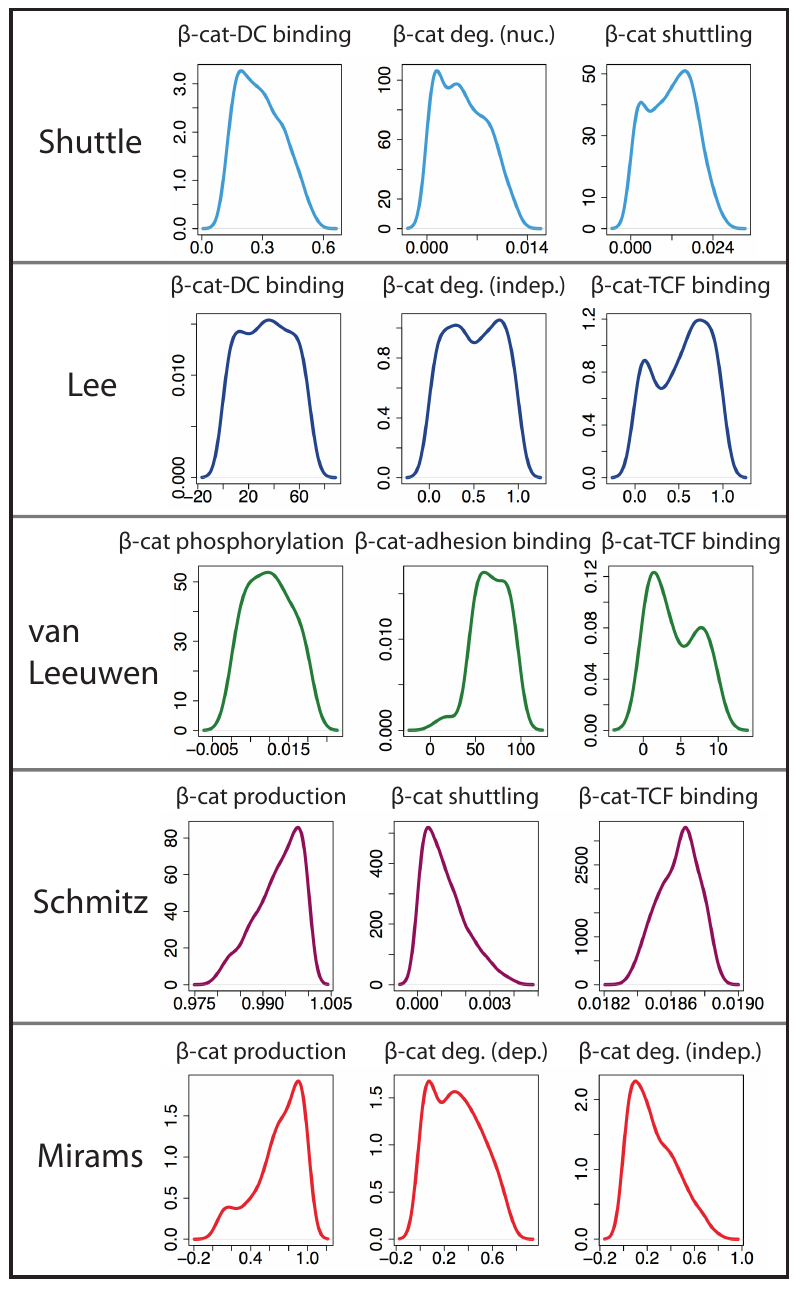}
\caption{Marginal posterior distributions for the parameter fit for each model under investigation. Each plot shows the frequency of parameter values for a given parameter of a model. dep./(indep.) degradation refers to degradation that is/(is not) influenced by Wnt signaling.}
\label{fig-posts}
\end{center}
\end{figure}

For the shuttle model, the parameters studied are the rate of binding of \bcat and TCF, the rate of degradation of \bcat in the nucleus, and the \bcat shuttling rate between cytoplasm and nucleus. The prior used for each parameter is [0,1]. From the results we see that the degradation and shuttling parameters are narrowly defined, whereas the rate of \bcat binding to DC spans a broader range of values. 
\par
For the Lee {\em et al.} model, we study the $\beta$-catenin-DC binding rate, that has a prior of $[0,100]$, the \bcat degradation rate independent of the DC, and the rate of binding of \bcat and TCF. The latter two parameters both have priors of $[0,1]$. The posterior distribution shows that the $\beta$-catenin-DC binding parameter takes values over the lower half of its prior range, whereas the other two parameters can take any values spanning the prior range. We deduce that for this model the parameter that has the greatest impact on outcome is the  $\beta$-catenin-DC binding rate.
\par
For the van Leeuwen {\em et al.} model, we study  $\beta$-catenin phosphorylation (leading to degradation), $\beta$-catenin binding to adhesion molecules, and $\beta$-catenin-TCF binding. The priors over these parameters are $[0,10]$, $[0,100]$ and $[0,10]$, respectively. One of these  parameters exerted significant control over outcome: the $\beta$-catenin phosphorylation rate, which must be low in order  to fit the \bcat increase in response to Wnt stimulus that the data describe. The parameters controlling \bcat binding to adhesion molecules and \bcat binding to TCF can vary over a much wider range of values.
\par
For the Schmitz {\em et al.} model, we study the \bcat production rate, the \bcat shuttling rate, and the rate of binding of \bcat to TCF. The prior used for each of the parameters is $[0,1]$ and we see that the marginal posterior distributions are relatively stiff: each parameter is constrained to lie within a narrow range relative to its prior. The rates of \bcat shuttling and binding to TCF are required be low, whilst the rate of \bcat production is required to be high in order to fit the data here.
\par
For the Mirams {\em et al.} model, which is specified by a single equation that describes the change in \bcat concentration over time, we study the \bcat production rate, and two \bcat degradation rates: dependent on and independent of the DC. The prior used for each of the parameters is $[0,1]$. The marginal posteriors are moderately stiff, that is, they each extend over the prior range, but within this range each shows a preference for higher or lower values. Both \bcat degradation rates take lower values; and the rate of production of \bcat takes higher values, in order to provide a good fit to the data.
\par
This analysis has described in some detail how the fits in Figure 3C were obtained, and what information can be gained from studying them. As we discussed in the main text, the data here are not of sufficient quality to choose between models. Since we have fixed parameters at possibly unrealistic values (due to different species parameter estimates from the literature) and we can still fit the free parameters, our inference demonstrates the relative simplicity of the data in comparison to the models; therefore we do not ascribe great weight to analysis of the posterior distributions. 


\section{Extended methods and analysis II: Multistationarity}

In this section, we prove the statements in the main text relating to the preclusion and assertion of monostationarity, multistationarity, and multistability of the different models of Wnt signalling.

\subsection{Qualitative approaches for preclusion/ determination of multistationarity}

Theorems developed in chemical reaction network theory (CRNT) enable us to relate qualitative dynamical behavior with the structure of the chemical reactions without relying on specific values of the parameters. Since we are interested in biological systems, we consider the behavior of the models in the positive orthant. We note that some notions 
(e.g., multistability) differ within the CRNT literature so we define them here:

\begin{defn} Consider a mathematical model that comprises a system of time-dependent ordinary differential equations $x'=F(x,\kappa)$ where $x$ are the dependent variables, $F$ defines the system kinetics and the constant coefficients $\kappa$ are the system parameters, A {\em steady state solution} of this mode satisfies $F(x,\kappa)=0$.\end{defn}
\begin{defn} A model is {\em injective} when $F(x,\kappa) = F(\tilde{x},\kappa)$ implies $x= \tilde{x}$. \end{defn}
\begin{remark} In the context of biological systems, we are interested only in injectivity for $x\in \mathbb{R}^+_{>0}$. Note that an injective model cannot have multiple positive steady states. \end{remark}

\begin{defn} A model has the capacity for {\em multistationarity} if it has multiple biologically feasible (i.e., positive) steady state solutions, for some values of the parameters $\kappa$ and total concentration amounts.A model exhibits  {\em monostationarity} when it has at most one steady state. \end{defn}
\begin{defn} A steady state solution is {\em stable} if it is locally asymptotically stable, i.e., if the real part of the eigenvalues of the associated Jacobian matrix, evaluated at the steady state, are all negative.\end{defn}
\begin{defn} A model has the capacity for {\em multistability} if two or more of the positive steady states are stable, for some value of parameters and total concentration amounts.\end{defn}
We are interested in whether these multiple positive states are stable (i.e., accessible), which is of particular biological importance for cellular decision making. If a system has two positive steady states, and only one is ever stable, the system cannot choose between states, for example, cell fate.

\par
The field of CRNT initially focused on a specific structural property of a network called deficiency, which could preclude multistability \cite{Feinberg:1987tk,Feinberg:1988ei}. Then theorems were proved for  
precluding/asserting multiple equilibria by studying the cycles in the graph of a network, or the sign of the determinant of the Jacobian; some of these approaches can provide conditions on the parameters for behaviors such as bistability and oscillations \cite{Craciun:2005ub,Craciun:2006et,Craciun:2006ki,Feliu:2011ur,Feliu:2013dz,craciun:pnasu:2006}.
Methods for precluding and asserting multistationarity may extend beyond chemical reaction networks \cite{Muller:2013vr, Feliu:2014vi, Lipson:2007vg}.  An excellent survey that reviews and presents a comprehensive techniques for multistationarity was recently written by Joshi and Shiu \cite{Shiu:2014}. We have applied tests and developed algorithms for chemical reaction networks to our Wnt signaling models.

\subsection{Details of multistationarity analysis}
We start by testing each model for injectivity, noting that while injectivity precludes multistationarity, failure of injectivity does not imply multistationarity. We use the algorithms in the CRNT Toolbox to determine whether the system can ever admit multiple positive steady states--multistationarity \cite{CRNT-toolbox}. Other methods have recently been developed that rely on different algorithms that have been reported to handle larger systems (tens to hundreds of reactions)  \cite{Feliu:2013dz,GraTeLPygraphtheo:2014gp}.   We present the results on the models, starting first with those that can only admit at most one state and those that can admit multiple.


\vspace{3mm}
\noindent {\bf van Leeuwen \etal model:}
This model is injective and hence does not admit multiple positive steady states for any value of parameters or total concentration amounts.

\vspace{3mm}
\noindent {\bf Lee \etal model:}
This model fails injectivity, but it cannot admit multiple positive steady states for any values of the system parameters and/or total concentration amounts (algorithms within \cite{CRNT-toolbox}).


\vspace{3mm}
\noindent {\bf Mirams \etal model:} 
This model is deficiency 1 in chemical reaction network theory. By running algorithms in the CRNT toolbox, it cannot admit multiple positive steady states, regardless of the values that the model parameters take.

\vspace{3mm}
\noindent {\bf Schmitz \etal model:} 
We find that this model fails injectivity, and admits two positive steady states. The CRNT Toolbox provides a sample parameter set that yields two steady states, one stable and one unstable. 
In this case, the system is sufficiently simple that we may derive analytical expressions for the steady state solutions: setting $\frac{d}{dt}=0$ in the Schmitz \etal model ODEs and manipulating the resulting algebraic equations supplies the following expressions for $Y_{an}, Y_i, X_p, X_{pn}, C_{XY}, C_{XYn}, T$ and $C_{XT}$ in terms of $X, X_n$ and $Y_a$:
\[ 
Y_{an} = \frac{\delta_3}{\delta_4)}Y_a, \;
Y_i = \frac{\delta_{15}}{\delta_{16}} Y_a, \; 
X_p = \frac{\delta_7}{\delta_{13}} \: \frac{\delta_5}{\delta_6+\delta_7} XY_a, \;
X_{pn} = \frac{\delta_8}{\delta_{14}} \: \frac{\delta_{10}}{\delta_9+\delta_{10}} XY_a, \]
\[
C_{XY} = \frac{\delta_5}{\delta_6+\delta_7} XY_a, \;
C_{XYn} = \frac{\delta_3}{\delta_4} \: \frac{\delta_8}{\delta_9+\delta_{10}} X_n Y_a, \]
\[ T = \left ( 1 + \frac{\delta_{11}}{\delta_{12}} X_n \right )^{-1} \; T_{TOT}, \;
C_{XT} = \frac{\delta_{11}}{\delta_{12}} \left ( 1 + \frac{\delta_{11}}{\delta_{12}} X_n \right )^{-1} \; X_n T_{TOT}.
\]
This requires that $Y_a = Y_a(X,X_n)$ satisfies

\[ Y_{TOT} = Y_a \left ( 1 + \frac{\delta_3}{\delta_4} + \frac{\delta_{15}}{\delta_{16}} +
\frac{\delta_5}{\delta_6 + \delta_7} X +  \frac{\delta_3}{\delta_4}\frac{\delta_8}{\delta_9+\delta_{10}} X_n \right ),\]
while $X_n$  depends linearly on $X$ via

\[ \left ( 1 + \frac{\delta_3}{\delta_4} + \frac{\delta_{15}}{\delta_{16}} \right ) 
= \frac{\delta_5}{\delta_6 + \delta_7}
\left ( \frac{\delta_7}{\delta_{0}} Y_{TOT} - 1\right ) X + \frac{\delta_3}{\delta_4} \frac{\delta_8}{\delta_9 + \delta_{10}}
\left ( \frac{\delta_{10}}{\delta_{0}} Y_{TOT} - 1 \right ) X_n,  \label{eq:star} \tag{$\star$} \]
and $X$ solves a quadratic of the form

\[ 0 = A X^2 + B X + C  \label{eq:diamond} \tag{$\diamond$}  \]
wherein the constant coefficients $A, B$ and $C$ are defined in terms of the model parameters. 
\par
The steady states are determined where Eq.~\eqref{eq:star} and Eq.~\eqref{eq:diamond} intersect. For physically realistic solutions, we require $X,X_n>0$ at intersections. Further, at most two points of intersection between parabola (Eq.~\ref{eq:diamond}) and straight line (Eq.~\ref{eq:star}) exist, which include a {\em linear} and {\em quadratic} term. Obviously the quadratic term only permits one positive steady state for positive rate constants. Therefore, we conclude that this model has at most two steady states. Since two stable states must be separated by an unstable state, and three do not exist, this model, as the CRNT toolbox suggests can admit at most one stable and one unstable.

\vspace{3mm}
\noindent {\bf Shuttle model:}
The model fails injectivity and it is also possible to admit two positive steady states. The CRNT Toolbox provides a sample parameter set that yields the two steady states, both of which are stable.  As described in the main text, we analyzed how different species shuttling can affect the multistationarity by setting certain shuttling parameters to zero. For example, if no species shuttle or if \bcat shuttle, the model is injective and cannot ever admit multiple positive steady states. As additional sets species shuttling between spatial compartments are analyzed, the model fails injectivity. For example, shuttling of only the Dsh and \bcat, or only open destruction complex and \bcat fails injectivity but cannot admit multiple positive steady states. However, if \bcat and closed destruction complex can shuttle, the system is capable of multiple steady states; however we find the existence of multiple stable steady states when \bcat, closed destruction complex and Dsh can shuttle.

\begin{table}
\begin{center}
\begin{tabular}{|c|c|c|c|c|c|c|c|}\hline
	 \multicolumn{8}{|c|}{Species that are allowed to shuttle}  \\ \hline
	\bcat & & \y & \y & \y &  \y&  \y&  \y  \\
	DC$_c$ & &  & \y &  &  & \y &  \y   \\
	Dsh & & &  & \y &  &  \y&  \y \\
	DC$_o$ & &  &  &  & \y &  &  \y   \\ \hline
	Injective &  Yes & Yes & No  &  No &  No &  No &  No \\
	Multi & No & No & Yes &  No & No  &  Yes&  Yes \\ 
	\# stable & &s & 1 & & & 2  & 2  \\ \hline
						\end{tabular}
\caption{Test performed to determine which species are required to shuttle for multistationarity. A positive injective test rules out multistationarity.}
\label{tab-injective}
\end{center}
\end{table}

%

\subsubsection{Conditions for monostationarity/multistationarity for the shuttle model}
We use the Jacobian injectivity approach as described in \cite{Harrington:2013gp} that has been implemented in Mathematica, which may provide sufficient conditions for monostationarity and necessary conditions for multistationarity. The coefficients of the determinant of the Jacobian must have the same sign for the model to be injective and hence, assert monostationarity (see \cite{Feliu:2014vi, Muller:2013vr, Feliu:2013dz, Feliu:2012ft}). We are able provide sufficient conditions for monostationarity by determining when all the coefficients of the determinant of the Jacobian to have the same sign.  There are 108 coefficients of the full shuttle model with 3 shuttling species, of which all but 24 coefficients had negative sign. We determined it was possible for all the positive coefficients of the determinant of the Jacobian to all have negative sign if the following conditions held (see non-repeating sufficient conditions for monostationarity in table). 

From these, we can determine necessary conditions for multistationarity to occur by negating coefficients of monostationarity conditions, i.e., flip the inequality sign.  The two shortest necessary conditions for multistationarity are C2 or C3:
$k_3 k_{15} + k_3 k_{24} - k_{14} k_{24}>0$
or
$k_5 k_{14} - k_3 k_{25} + k_{14} k_{25}>0;$
and by rearranging, we find that these are dependent on either the shuttling of \bcat or the degradation rates of \bcat
\begin{align*}
\underbrace{\dfrac{k_{3} -k_{14}}{k_{3}k_{15}} }_{\text{degradation rates}}> \underbrace{-\dfrac{1}{k_{24}} }_{{\text{shuttling rates}}} \quad
 \text{or}
 \underbrace{ \dfrac{k_{3} -k_{14}}{k_{5}k_{14}} }_{\text{degradation rates}}< \underbrace{\dfrac{1}{k_{25}} }_{{\text{shuttling rates}}}.
  \end{align*}

\begin{table}[htb!]
{\tiny
\begin{tabular}{|c | C{12.5cm}|} \hline
C1 		& $- k_{1} k_{6} k_{9} k_{16} k_{18} (k_{20} + k_{21}) k_{23} (k_{3} k_{11} k_{13} k_{15} k_{26} k_{28} k_{30} + k_{3} k_{11} k_{14} k_{15} k_{26} k_{28} k_{30} + k_{3} k_{11} k_{13} k_{24} k_{26} k_{28} k_{30} +     k_{3} k_{11} k_{14} k_{24} k_{26} k_{28} k_{30} + k_{3} k_{11} k_{12} k_{15} k_{26} k_{28} k_{31} + k_{3} k_{11} k_{12} k_{24} k_{26} k_{28} k_{31} - k_{11} k_{12} k_{14} k_{24} k_{26} k_{28} k_{31})  <0$ 	\\ \hline
C2         & $k_3 k_{15} + k_3 k_{24} - k_{14} k_{24}<0$ \\ \hline
C3         & $k_5 k_{14} - k_3 k_{25} + k_{14} k_{25}<0$ \\ \hline
C5 		& $- k_{1} k_{6} k_{9} k_{16} k_{18} k_{19} k_{23} (k_{3} k_{11} k_{13} k_{15} k_{26} k_{28} k_{30} + k_{3} k_{11} k_{14} k_{15} k_{26} k_{28} k_{30} +  k_{3} k_{11} k_{13} k_{24} k_{26} k_{28} k_{30} +  k_{3} k_{11} k_{14} k_{24} k_{26} k_{28} k_{30} +  k_{3} k_{11} k_{12} k_{15} k_{26} k_{28} k_{31} +  k_{3} k_{11} k_{12} k_{24} k_{26} k_{28} k_{31} -  k_{11} k_{12} k_{14} k_{24} k_{26} k_{28} k_{31}) < 0$	\\ \hline
C8		& $- k_{1} k_{6} k_{8} k_{9} k_{19} k_{22} (k_{3} k_{13} k_{15} k_{16} k_{21} k_{26} k_{28} k_{30} + k_{3} k_{14} k_{15} k_{16} k_{21} k_{26} k_{28} k_{30} +  k_{3} k_{12} k_{14} k_{17} k_{21} k_{26} k_{28} k_{30} +  k_{3} k_{12} k_{14} k_{18} k_{21} k_{26} k_{28} k_{30} +  k_{3} k_{13} k_{16} k_{21} k_{24} k_{26} k_{28} k_{30} +  k_{3} k_{14} k_{16} k_{21} k_{24} k_{26} k_{28} k_{30} +  k_{3} k_{12} k_{14} k_{17} k_{21} k_{27} k_{28} k_{30} +  k_{3} k_{12} k_{14} k_{18} k_{21} k_{27} k_{28} k_{30} +  k_{3} k_{12} k_{14} k_{17} k_{21} k_{27} k_{29} k_{30} + k_{3} k_{12} k_{14} k_{18} k_{21} k_{27} k_{29} k_{30} + k_{3} k_{12} k_{15} k_{16} k_{21} k_{26} k_{28} k_{31} + k_{3} k_{12} k_{16} k_{21} k_{24} k_{26} k_{28} k_{31} - k_{12} k_{14} k_{16} k_{21} k_{24} k_{26} k_{28} k_{31}) < 0$		\\ \hline
C12 		& $ - k_{1} k_{6} k_{8} k_{9} k_{19} k_{22} (k_{3} k_{13} k_{15} k_{17} k_{21} k_{26} k_{28} k_{30} +  k_{3} k_{14} k_{15} k_{17} k_{21} k_{26} k_{28} k_{30} +  k_{3} k_{13} k_{15} k_{18} k_{21} k_{26} k_{28} k_{30} +  k_{3} k_{14} k_{15} k_{18} k_{21} k_{26} k_{28} k_{30} +  k_{3} k_{13} k_{17} k_{21} k_{24} k_{26} k_{28} k_{30} +  k_{3} k_{14} k_{17} k_{21} k_{24} k_{26} k_{28} k_{30} +  k_{3} k_{13} k_{18} k_{21} k_{24} k_{26} k_{28} k_{30} +  k_{3} k_{14} k_{18} k_{21} k_{24} k_{26} k_{28} k_{30} +  k_{3} k_{13} k_{15} k_{17} k_{21} k_{27} k_{28} k_{30} +  k_{3} k_{14} k_{15} k_{17} k_{21} k_{27} k_{28} k_{30} +  k_{3} k_{13} k_{15} k_{18} k_{21} k_{27} k_{28} k_{30} +  k_{3} k_{14} k_{15} k_{18} k_{21} k_{27} k_{28} k_{30} +  k_{3} k_{13} k_{17} k_{21} k_{24} k_{27} k_{28} k_{30} +  k_{3} k_{14} k_{17} k_{21} k_{24} k_{27} k_{28} k_{30} +  k_{3} k_{13} k_{18} k_{21} k_{24} k_{27} k_{28} k_{30} +  k_{3} k_{14} k_{18} k_{21} k_{24} k_{27} k_{28} k_{30} +  k_{3} k_{13} k_{15} k_{17} k_{21} k_{27} k_{29} k_{30} +  k_{3} k_{14} k_{15} k_{17} k_{21} k_{27} k_{29} k_{30} +  k_{3} k_{13} k_{15} k_{18} k_{21} k_{27} k_{29} k_{30} +  k_{3} k_{14} k_{15} k_{18} k_{21} k_{27} k_{29} k_{30} +  k_{3} k_{13} k_{17} k_{21} k_{24} k_{27} k_{29} k_{30} +  k_{3} k_{14} k_{17} k_{21} k_{24} k_{27} k_{29} k_{30} +  k_{3} k_{13} k_{18} k_{21} k_{24} k_{27} k_{29} k_{30} +  k_{3} k_{14} k_{18} k_{21} k_{24} k_{27} k_{29} k_{30} +  k_{3} k_{12} k_{15} k_{17} k_{21} k_{26} k_{28} k_{31} +  k_{3} k_{12} k_{15} k_{18} k_{21} k_{26} k_{28} k_{31} +  k_{3} k_{12} k_{17} k_{21} k_{24} k_{26} k_{28} k_{31} -  k_{12} k_{14} k_{17} k_{21} k_{24} k_{26} k_{28} k_{31} +  k_{3} k_{12} k_{18} k_{21} k_{24} k_{26} k_{28} k_{31} -  k_{12} k_{14} k_{18} k_{21} k_{24} k_{26} k_{28} k_{31} +  k_{3} k_{12} k_{15} k_{17} k_{21} k_{27} k_{28} k_{31} + k_{3} k_{12} k_{15} k_{18} k_{21} k_{27} k_{28} k_{31} +  k_{3} k_{12} k_{17} k_{21} k_{24} k_{27} k_{28} k_{31} -  k_{12} k_{14} k_{17} k_{21} k_{24} k_{27} k_{28} k_{31} +  k_{3} k_{12} k_{18} k_{21} k_{24} k_{27} k_{28} k_{31} -  k_{12} k_{14} k_{18} k_{21} k_{24} k_{27} k_{28} k_{31} +  k_{3} k_{12} k_{15} k_{17} k_{21} k_{27} k_{29} k_{31} + k_{3} k_{12} k_{15} k_{18} k_{21} k_{27} k_{29} k_{31} + k_{3} k_{12} k_{17} k_{21} k_{24} k_{27} k_{29} k_{31} - k_{12} k_{14} k_{17} k_{21} k_{24} k_{27} k_{29} k_{31} +  k_{3} k_{12} k_{18} k_{21} k_{24} k_{27} k_{29} k_{31} -  k_{12} k_{14} k_{18} k_{21} k_{24} k_{27} k_{29} k_{31}) < 0$
		\\ \hline
C14 		& $- k_{1} k_{6} k_{8} (k_{10} + k_{11}) k_{19} k_{22} (k_{3} k_{13} k_{15} k_{16} k_{21} k_{26} k_{28} k_{30} + k_{3} k_{14} k_{15} k_{16} k_{21} k_{26} k_{28} k_{30} + k_{3} k_{12} k_{14} k_{17} k_{21} k_{26} k_{28} k_{30} + k_{3} k_{12} k_{14} k_{18} k_{21} k_{26} k_{28} k_{30} + k_{3} k_{13} k_{16} k_{21} k_{24} k_{26} k_{28} k_{30} + k_{3} k_{14} k_{16} k_{21} k_{24} k_{26} k_{28} k_{30} +  k_{3} k_{12} k_{14} k_{17} k_{21} k_{27} k_{28} k_{30} + k_{3} k_{12} k_{14} k_{18} k_{21} k_{27} k_{28} k_{30} + k_{3} k_{12} k_{14} k_{17} k_{21} k_{27} k_{29} k_{30} + k_{3} k_{12} k_{14} k_{18} k_{21} k_{27} k_{29} k_{30} +  k_{3} k_{12} k_{15} k_{16} k_{21} k_{26} k_{28} k_{31} +  k_{3} k_{12} k_{16} k_{21} k_{24} k_{26} k_{28} k_{31} -  k_{12} k_{14} k_{16} k_{21} k_{24} k_{26} k_{28} k_{31}) < 0$		\\ \hline
C18 		& $- k_{1} k_{6} k_{8} (k_{10} + k_{11}) k_{19} k_{22} (k_{3} k_{13} k_{15} k_{17} k_{21} k_{26} k_{28} k_{30} + k_{3} k_{14} k_{15} k_{17} k_{21} k_{26} k_{28} k_{30} + k_{3} k_{13} k_{15} k_{18} k_{21} k_{26} k_{28} k_{30} + k_{3} k_{14} k_{15} k_{18} k_{21} k_{26} k_{28} k_{30} + k_{3} k_{13} k_{17} k_{21} k_{24} k_{26} k_{28} k_{30} + k_{3} k_{14} k_{17} k_{21} k_{24} k_{26} k_{28} k_{30} + k_{3} k_{13} k_{18} k_{21} k_{24} k_{26} k_{28} k_{30} + k_{3} k_{14} k_{18} k_{21} k_{24} k_{26} k_{28} k_{30} + k_{3} k_{13} k_{15} k_{17} k_{21} k_{27} k_{28} k_{30} + k_{3} k_{14} k_{15} k_{17} k_{21} k_{27} k_{28} k_{30} + k_{3} k_{13} k_{15} k_{18} k_{21} k_{27} k_{28} k_{30} + k_{3} k_{14} k_{15} k_{18} k_{21} k_{27} k_{28} k_{30} + k_{3} k_{13} k_{17} k_{21} k_{24} k_{27} k_{28} k_{30} + k_{3} k_{14} k_{17} k_{21} k_{24} k_{27} k_{28} k_{30} + k_{3} k_{13} k_{18} k_{21} k_{24} k_{27} k_{28} k_{30} +k_{3} k_{14} k_{18} k_{21} k_{24} k_{27} k_{28} k_{30} +k_{3} k_{13} k_{15} k_{17} k_{21} k_{27} k_{29} k_{30} +k_{3} k_{14} k_{15} k_{17} k_{21} k_{27} k_{29} k_{30} +k_{3} k_{13} k_{15} k_{18} k_{21} k_{27} k_{29} k_{30} +k_{3} k_{14} k_{15} k_{18} k_{21} k_{27} k_{29} k_{30} +k_{3} k_{13} k_{17} k_{21} k_{24} k_{27} k_{29} k_{30} +k_{3} k_{14} k_{17} k_{21} k_{24} k_{27} k_{29} k_{30} +k_{3} k_{13} k_{18} k_{21} k_{24} k_{27} k_{29} k_{30} +k_{3} k_{14} k_{18} k_{21} k_{24} k_{27} k_{29} k_{30} +k_{3} k_{12} k_{15} k_{17} k_{21} k_{26} k_{28} k_{31} +k_{3} k_{12} k_{15} k_{18} k_{21} k_{26} k_{28} k_{31} +k_{3} k_{12} k_{17} k_{21} k_{24} k_{26} k_{28} k_{31} - k_{12} k_{14} k_{17} k_{21} k_{24} k_{26} k_{28} k_{31} + k_{3} k_{12} k_{18} k_{21} k_{24} k_{26} k_{28} k_{31} - k_{12} k_{14} k_{18} k_{21} k_{24} k_{26} k_{28} k_{31} + k_{3} k_{12} k_{15} k_{17} k_{21} k_{27} k_{28} k_{31} + k_{3} k_{12} k_{15} k_{18} k_{21} k_{27} k_{28} k_{31} + k_{3} k_{12} k_{17} k_{21} k_{24} k_{27} k_{28} k_{31} - k_{12} k_{14} k_{17} k_{21} k_{24} k_{27} k_{28} k_{31} + k_{3} k_{12} k_{18} k_{21} k_{24} k_{27} k_{28} k_{31} - k_{12} k_{14} k_{18} k_{21} k_{24} k_{27} k_{28} k_{31} + k_{3} k_{12} k_{15} k_{17} k_{21} k_{27} k_{29} k_{31} + k_{3} k_{12} k_{15} k_{18} k_{21} k_{27} k_{29} k_{31} + k_{3} k_{12} k_{17} k_{21} k_{24} k_{27} k_{29} k_{31} - k_{12} k_{14} k_{17} k_{21} k_{24} k_{27} k_{29} k_{31} +k_{3} k_{12} k_{18} k_{21} k_{24} k_{27} k_{29} k_{31} - k_{12} k_{14} k_{18} k_{21} k_{24} k_{27} k_{29} k_{31}) < 0$		\\ \hline
\end{tabular}}
\end{table}

\clearpage
\subsection{Bifurcation analysis}
We analyzed the shuttle model in a bistable parameter regime using the parameters given from the CRNT Toolbox: \\

\begin{tabular}{llllll}
$k_1=92.331732$ & $k_2=0.86466471$ & $k_3=79.9512906$ &$k_4=97.932525$ &$k_5=1$ \\
$k_6=3.4134082$ &$k_7=0.61409879$& $k_8=0.61409879$& $k_9=3.4134082$& $k_{10}=0.98168436$\\ 
$k_{11}=0.98168436$& $k_{12}=4.7267833$ &$k_{13}=0.17182818$& $k_{14}=0.68292191$& $k_{15}=1$\\
 $k_{16}=3.2654672$& $k_{17}=0.61699064$& $k_{18}=0.61699064 $&$k_{19}=37.913879$& $k_{20}=0.86466471$\\ 
 $k_{21}=0.86466471$& $k_{22}=0.99326205$& $k_{23}=0.99326205$& $k_{24}=1$ &$k_{25}=5.9744464$\\
  $k_{26}=1.7182818$& $k_{27}=1.7182818$& $k_{28}=1.7182818$& $k_{29}=1.7182818$& $k_{30}=0.55950727$\\ 
$k_{31}=1.0117639$ & & & & &\\
\end{tabular}\\

\noindent and total amounts 
$DC_{\rm TOT} = 16.4734,$  $Dsh_{\rm TOT}=4.9951,$ $P_{\rm TOT}=1.60063,$ $Pn_{\rm TOT}=1.20891,$ and $T_{\rm TOT}=2.77566.$
While these parameters are not biologically informed, the rate constants provide an opportunity to study the qualitatively behavior the model. We do not perform an exhaustive search, instead we vary the parameters that are involved in the repeating necessary conditions for multistationarity. By varying on parameter (either the degradation or shuttle rates), we notice that the system exhibits a hysteresis or memory, i.e., the threshold value of the parameter to switch from a low to high state is not the same value to switch from a high to low state (see Fig 2 in the main text). This sample parameter value demonstrates the capacity for a bistable switch, which is a common phenomena in cell fate switching mechanisms such as apoptosis  \cite{Ho:2010hb,scott:n:2009,Li:2000fk, bagci:bj:2006}. We do not make any predictions about how the behavior of the species are affected (as well as the reversible/irreversible nature of the bifurcations) due to the large uncertainty in parameter values. All bifurcations are computed using {\tt Oscill8} (Available at: oscill8.sourceforge.net/doc) and visualized using {\tt MATLAB} (R2013a; The MathWorks, Natick, MA).

\section{Extended methods and analysis III: Matroids}
\subsection{Qualitative analysis of models with algebraic matroids}
When the solution set to a system of polynomial equations is an irreducible variety, the associated algebraic matroid encodes the dependencies among the variables. For a steady-state solution to a system of ODE's, the derivatives are set to $0$ and what remains is a system of polynomial equations among the species concentrations. These polynomials generate a \emph{polynomial ideal}, which has \emph{associated prime ideals}. The irreducible components of the solution set correspond to the associated primes. Our approach is to compute the matroid associated to each irreducible component.  This calculation is carried out using elimination ideals in the computer algebra software {\tt Macaulay2} (Available at http://www.math.uiuc.edu/Macaulay2). The commands used to do this can be found at {http://math.berkeley.edu/$\sim$zhrosen/matroids.html} and Wnt specific code is available at {http://math.berkeley.edu/~zhrosen/wntCode.html}.

Explicitly, Macaulay2 performs the following computations:

\vspace{5mm}

{\bf Algorithm 1.} (For computation of bases.)

Let $E$ be the set of coordinates and let $I$ be the steady-state ideal.
\begin{enumerate}
\item Compute the dimension of the ideal (command: {\tt dim I}). This is the rank of the matroid $r$.
\item Enumerate all subsets of variables of size $r$. 
\item For each subset $S$, eliminate the other variables in the ideal (command: {\tt eliminate(I,E-S)}).
\item If the resulting ideal is the zero ideal (Macaulay output: {\tt ideal()}), then return ``base". Otherwise, return ``not a base."
\end{enumerate}

\pagebreak
{\bf Algorithm 2.} (For computation of circuits.)

Let $E$ be the set of coordinates and let $I$ be the steady-state ideal.
\begin{enumerate}
\item Again compute the dimension of the ideal (command: {\tt dim I}). This is the rank of the matroid $r$.
\item Enumerate all subsets of variables of size less than or equal to $r+1$. 
\item For each subset $S$, eliminate the other variables in the ideal (command: {\tt J = eliminate(I,E-S)}).
\item If the resulting ideal is principal and the generator uses all variables in $S$ (command: {\tt codim J == 1 and support((flatten entries gens J)\#0) == S}), then return ``circuit". Otherwise, return ``not a circuit."
\end{enumerate}

Moving from a set of bases or circuits to an affine representation is done by hand.

This technique for analyzing a CRN model has several key advantages: \begin{enumerate}

\item \emph{Parameters}: The algebraic matroid makes no numerical computations as to actual concentrations; it only encodes the existence or absence of a polynomial relationship among a set of species. For this reason, parameter values, if chosen generically, do not affect the matroid. In a model where rate parameters are unknown, this is very useful, since random real numbers can be substituted.

\item \emph{Linearity}: In the systems analyzed here, Gr\"{o}bner basis computations were quick in outputting the matroid. However, in larger systems, computational complexity makes Gr\"{o}bner basis computation unrealistic. Algebraic matroids over the real numbers (which we are dealing with) can be linearized by passing to differentials. The algebraic structure can therefore be easily computed using linear algebra.

\item \emph{Global Structure}: Elimination ideals have been used to analyze CRNs in the past, but the advantage of using matroids is that the entire structure of algebraic dependencies is represented, not only chosen subsets.

\end{enumerate}

\subsection{Details of matroid analysis} 

The matroids for the various models were included in Figure 4 of the main text. Here, we include the algebraic information that led to the output of these images.

\vspace{3mm}
\noindent {\bf van Leeuwen \etal model:}
This model includes ODEs with rational functions. To calculate the ideal associated to the steady state, we clear all denominators. The resulting ideal has two associated primes corresponding to: (1) a $2$-dimensional plane, and (2) a $0$-dimensional variety of degree $2$, which could correspond to at most two points.  The matroid depicted in the Figure is associated to Component (1). This ideal is defined by nine linear equations.

The matroid has five bases (independent sets of size $2$), each of which contain $X_p$; the matroid has five loops and ten circuits of size $2$. In analyzing the matroid, we keep in mind that the concentration of $X_p$ is independent of all other measurements, the concentration of loops is fixed by rate parameters alone, and the remaining species will all be determined by measuring just one of them.

\vspace{3mm}
\noindent {\bf Schmitz \etal model:} The steady-state solution set is a $2$-dimensional irreducible variety of degree $5$. The associated ideal can be generated by six linear polynomials, and three polynomials of degree $2$. The matroid has $19$ bases and $45$ circuits: $36$ circuits of size $2$, and $9$ circuits of size $3$.
\vspace{3mm}

\noindent {\bf Lee \etal model:} The steady-state solution set is a $3$-dimenisonal irreducible variety of degree $7$. The associated ideal can be generated by eight linear polynomials, and five polynomials of degree $2$.  The matroid has $62$ bases and $62$ circuits: $3$ loops, $16$ circuits of size $2$, $38$ of size 3, and $5$ circuits of size $4$.

\vspace{3mm}
\noindent {\bf Shuttle model:}  The steady-state solution set for the shuttle model has the most complicated matroid. The variety has two irreducible components of dimension $5$: (1) has degree $30$ and (2) has degree $6$. Comopnent (1) can be generated by seven linear polynomials, eight polynomials of degree $2$, and one of degree $3$. Component (2) can be generated by eleven linear polynomials, and three polynomials of degree $2$.

The matroid for component (1), $\mathcal{M}_1$, has no loops. It has $2,389$ bases (independent sets of size $5$). There are $951$ circuits (minimal dependent sets): six of size $2$, $41$ of size $3$, $269$ of size $4$, $505$ of size $5$, and $130$ of size $6$. The nontrivial rank 1 and 2 flats are as depicted in Figure 4. The remaining flats assume that the line placement is as generic as possible with one exception: The line formed by $\{C_{XY}, C_{XYn}, X,X_n,Y_a,Y_{an}\}$ and the line formed by $P$ and $P_n$ are coplanar.

\section{Extensions of parameter-free approaches for insight into models}
When data from a model clearly supports a specific behavior --- whether monostable, bistable, or oscillatory, the qualitative approaches such as those mentioned in the multistationarity section may be a good first step for classifying models, especially if data isn't of sufficient quality to estimate parameters. However if steady state data is available, then matroids may be helpful for guiding parameter-free model discrimination. 

\subsection{Steady state analysis and matroids}
For smaller models, such as the Schmitz \etal model, the steady states can be determined by solving Eq.~\ref{eq:diamond} and Eq.~\ref{eq:star}, which are functions of $X$ and $X_n$. Recall the steady state expressions are in terms of $X$ and $X_n$ because all the other variables were eliminated by using conservation laws and variable substitution. Effectively, we can solve for $X$ in Eqs.~\ref{eq:diamond} and \ref{eq:star} as a function of $X_n$ and in doing so we find that there is an algebraic dependence between $X$ and $X_n$. In fact, the matroid associated to the Schmitz \etal model highlights this dependence (see Figure 4 in main text) in the circuit $\{X,X_n\}$:  
\begin{align*}
\delta_0 \delta_3 \delta_4 \delta_6 (\delta_8+\delta_9)X^2 + (\delta_0 \delta_2 \delta_7 \delta_9(\delta_5+\delta_6)- \delta_1 \delta_3 \delta_4 \delta_6(\delta_8+\delta_9))X X_n - \delta_1 \delta_2 \delta_7 \delta_9(\delta_5+\delta_6) X_n^2
\end{align*}

\noindent When the steady states cannot be solved analytically, the matroid gives all possible dependencies in the form of a circuit polynomial, which may be helpful for deciding which variables to measure (which are dependent) in order to test model/data compatibility. 

\subsection{Using matroids for model discrimination}
Suppose $\beta$-catenin localization data (non-nucleus $X$ and nucleus $X_n$) exist as published in \cite{Tan:2014ii}. With observable measurements $X,X_n$ denoted by lower-case, and data by a hat: $\boldsymbol{\hat{x}_{\obs}} = \{\hat{x},\hat{x}_n\}$, we would like to determine whether we can rule out Wnt models using a parameter-free framework.
In particular, we seek to apply our matroid analysis to discriminate between models. First, we consult all models and find that only two models have nuclear and non-nuclear $\beta$-catenin variables: the Schmitz \etal model give a circuit polynomial relationship whereas the shuttle model $\beta$-catenin species are algebraically independent (i.e., the smallest circuit must have three species thus any data $(\hat{x},\hat{x}_n)$ will be compatible with the shuttle model).

The algebraic relationship described by the circuit polynomial is a type of Gr\"obner basis with elimination of all other species, effectively a steady-state invariant: 
\begin{equation}
\mathcal{I}(\boldsymbol{x_{\obs}},\boldsymbol{\delta})= \sum_{j = 1}^{n} h_{j} \left( \boldsymbol{\delta} \right) \prod_{k = 1}^{N_{\obs}} x_{k}^{t_{jk}} =  0,
\end{equation} \label{eq:schmitzinvar} where again $\boldsymbol{{x}_{\obs}}$ are observable variables and $\boldsymbol{\delta}$ is a vector of parameters. 

Following \cite{Harrington:2012us}, we can perform model discrimination using observables measurements $\boldsymbol{\hat{x}_{obs}}$ by rewriting re-write Eq.~\ref{eq:schmitzinvar} as $I(\boldsymbol{y};\boldsymbol{d}) = \sum_{j=1}^n d_jy_j$ where $y_{j} = \prod_{k = 1}^{N_{\obs}} x_{k}^{t_{jk}}$ and $d_{j} = h_{j} (\boldsymbol{\delta})$, with $\boldsymbol{y} = (y_{1}, \dots, y_{n})$ and 
$\boldsymbol{d} = (d_{1}, \dots, d_{n})$. For the Schmitz \etal model, $\boldsymbol{y}= \{x^2,xx_n,x_n^2\}$ and $\boldsymbol{d} = \{h_1(\delta), h_2(\delta), h_3(\delta) \}.$
\par
Let $\varphi$ be the map taking $\boldsymbol{x}_{\obs}$ to $\boldsymbol{y}$. Then compatibility implies that transformed variables $\boldsymbol{\hat{y}} = \varphi(\boldsymbol{\hat{x}_{\obs}}$) are points on a hyperplane with coordinates $y_j$ defined by coefficients $\boldsymbol{d}$. 

Let $\boldsymbol{Y} \in \mathbb{R}^{m \times n}$ be the matrix with each row $i$ the transformed variables $\hat{y}$ are evaluated at datum $\boldsymbol{x_{\obs,i}}.$ If there exists a nontrivial vector $\boldsymbol{d}$ that satisfies $\boldsymbol{{Y}}\boldsymbol{d}=\boldsymbol{0}$ (i.e., a vector $\boldsymbol{d}$ resides in the kernel of $\boldsymbol{Y}$, found by using the singular value decomposition $\boldsymbol{Y} = \boldsymbol{U \Sigma V^T}$), then the model is compatible with the data. The singular values $\sigma_i\geq0$ are obtained from the diagonal of $\boldsymbol{\Sigma}.$ The smallest singular value $\sigma_{\text{min}}$ bounds the norm $|| \boldsymbol{Yd}||$ for any $\boldsymbol{d} \neq 0$ via $ \sigma_{\text{min}} = \min_{\left\| \boldsymbol{d} \right\| = 1} \left\| \boldsymbol{Y} \boldsymbol{d} \right\|,$ thus for $\sigma_{\text{min}}>0$ the data are not coplanar. 

The case for noisy data, statistical cut-offs for model rejection have been developed and performed here following \cite{Harrington:2012us}. For a moment, suppose we know our measurement noise $\epsilon$. Recall that we assume our observed data $\boldsymbol{\hat{x}_{\obs}} = \boldsymbol{x_{\obs}} + \boldsymbol{x} \epsilon Z$ where $\epsilon$ is the measurement noise and $Z \sim \mathcal{N}(0,1)$, i.e., a standard normal random variable. Our invariant is given in terms of transformed variables, therefore, the propagation of error through $\boldsymbol{\hat{y}}$ can be determined by the perturbation equation: $$\boldsymbol{\hat{y}} = \boldsymbol{y} + \Delta \boldsymbol{y} = \varphi(\boldsymbol{x} + \Delta \boldsymbol{x}).$$  

By expanding to first order, the error $\Delta \boldsymbol{\hat{y}}  = \nabla \varphi (\boldsymbol{\hat{x}_{\obs}}) \Delta \boldsymbol{\hat{x}_{\obs}}$ where $(\nabla \varphi)$ is the Jacobian of $\phi$ with elements $(\nabla \varphi)_{ij}= \partial y_i / \partial x_j $ evaluated at $\boldsymbol{\hat{x_{\obs}}}$. Then by assuming $\hat{y}_i$ are coplanar with $\|d\|=1$, then perturbed by measurement noise in $\boldsymbol{x}$, $$\Delta \hat{y}_j =  \epsilon_j  Z, \quad \epsilon_j = \epsilon \left[ \sum_{k=1}^{N_\obs} \left( (\nabla \varphi)_{jk}\hat{x}_k \right)^{2} \right]^{1/2} .$$

For the Schmitz \etal model, the Jacobian,
\begin{align*} 
\nabla \varphi = 
 \left( \begin{array}{cc}
 \partial y_1/ \partial x_1 &
\partial y_1/ \partial x_2 \\
\partial y_2/ \partial x_1 &
\partial y_2/ \partial x_2 \\ 
\partial y_3/ \partial x_1 &
\partial y_3/ \partial x_2   
 \end{array} \right)
 =
 \left( \begin{array}{cc}
 2x & 0 \\
 x_n & x \\
 0 & 2x_n 
 \end{array} \right),
 \quad
 \begin{array}{c}
\epsilon_1= 2\epsilon \hat{x}^2 ~ \\
\epsilon_2= 2\epsilon \hat{x}\hat{x_n}  \\ 
\epsilon_3= 2\epsilon \hat{x_n}^2. \end{array}
 \end{align*}

We seek to understand $\boldsymbol{Yd}$ whose vectors are perturbed from zero to 

$$\sum_{j = 1}^n d_j \Delta \hat{y}_j = \left( \sum_{j = 1}^n d_j^2 \epsilon_j^2 \right)^{1/2} Z = \| \boldsymbol{\epsilon} \| Z$$
using properties of random variables, where $\boldsymbol{\epsilon} = (\epsilon_{1}, \dots, \epsilon_{n})$ is a vector for each transformed datum $\boldsymbol{\hat{y}}.$ Since $\| d \| = 1$ the each row of $\boldsymbol{Y}$ must be rescaled by its effective error 
$$\epsilon_{\text{eff}} = \| \epsilon \| \geq \left( \sum_{j = 1}^{n} d_{j}^{2} \epsilon_{j}^{2} \right)^{1/2} $$
to give $\boldsymbol{Y'}$ and each entry of $\boldsymbol{Y'd}$ has the form $\tau_i^2 Z$ with variance $\tau_i^2  \leq 1$ for $i = 1, \ldots m.$ 

The coplanarity error is defined as
$$\Delta = \sigma_{\min}(\boldsymbol{Y'}) \leq \|Y'd\|,$$
which is bounded by the length of a normal random vector with variances $\tau_i^2 \leq 1$ (as described previously).  This distribution function is dominated by the length of a normal random vector with variances $\tau_i^2=1$, which is precisely the $\chi$ distribution with $m$ degrees of freedom. Statistically, that is,
 \begin{align}
 \Pr \left( \Delta \geq x \right) \leq \Pr \left( X \geq x \right), \quad X \sim \chi_{m},
\end{align}
so if $p_{\alpha}$ is the upper $\alpha$-percentile for $\chi_{m}$ then
\begin{align}
 \Pr \left( \Delta \geq p_{\alpha} \right) \leq \Pr \left( X \geq p_{\alpha} \right) = \alpha,
\end{align}
which gives a probability bound for any given threshold criterion for rejecting coplanarity.

If many steady state observations are available, for example, single cell concentration measurements of \bcat in the cytoplasm and nucleus, then these data can be substituted into the above, that is $\boldsymbol{{Y'}}\boldsymbol{d}=0,$ where $\boldsymbol{{Y'}}$ is the transformed data matrix. We assume we have such data through simulation. First we simulate data from the Schmitz \etal model and compare it to the Schmitz \etal steady-state invariant as a consistency check. Next we simulate data from the shuttle model and apply it to the Schmitz \etal invariant. In order to test compatibility, we simulated 100 cells from either the shuttle or the Schmitz \etal model, with multiplicative lognormal noise, and tested their compatibility with the Schmitz \etal model. We used a statistical cutoff at a 5\% significance level. Assuming $\alpha = 0.05$, the coplanarity cut-off is 11.15 for 100 data points. 

Finally we applied the method using time-course replicate experimental data of mammalian $\beta$-catenin localization as published in \cite{Tan:2014ii}. These three replicates of experimental $\beta$-catenin data, in the two compartments, are the minimal number of species and minimal number of replicates that can be used for the proposed matroid-informed coplanarity method. 

\begin{figure}[H]
\begin{center}
\includegraphics[width=\textwidth]{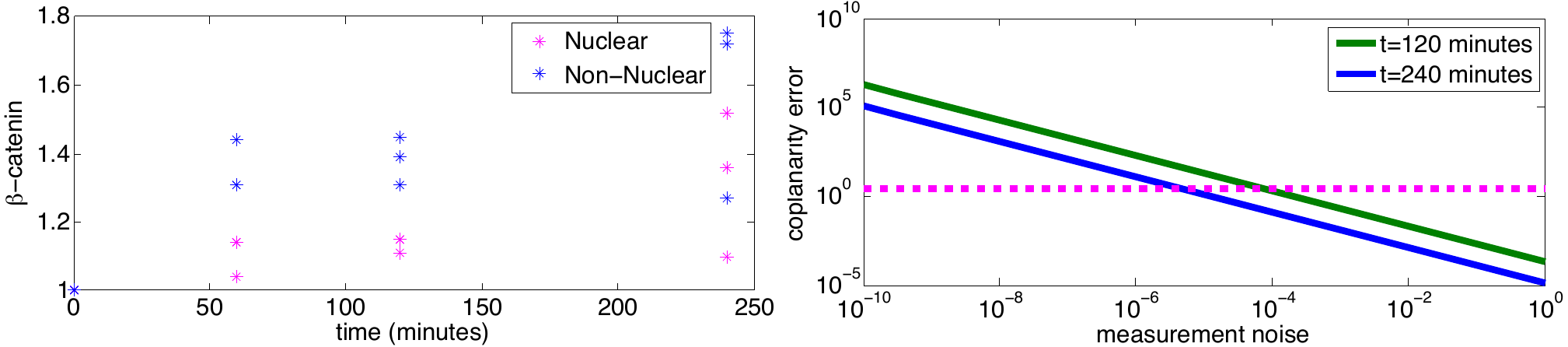}
\caption{Parameter-free model discrimination using mammalian \bcat data. (a) Reproduction of nuclear and non-nuclear \bcat data at four time points. (b) Coplanarity error of Schmitz model (null hypothesis) using 120 minute (green) and 240 minute (blue) replicate data from \cite{Tan:2014ii}. We find that for small noise, the Schmitz \etal model can be rejected when the coplanarity error is greater than the bound determined by $\chi_3$ (dashed magenta). }
\label{fig-posts}
\end{center}
\end{figure}

We used 120 minute (green) and 240 minute (blue) replicate data from \cite{Tan:2014ii} to calculate the coplanarity error of the Schmitz \etal model at different levels of measurement noise. This reaffirmed our conclusion that the Schmitz et al model can be rejected for small measurement noise (as we demonstrated using simulated data from the shuttle model).


\bibliographystyle{plain}
\bibliography{references,extrarefs}

\end{document}